\pdfoutput=1

\documentclass{aa}  
\usepackage{graphicx}
\usepackage{txfonts}
\usepackage[colorlinks=true,linkcolor=blue,allcolors=blue]{hyperref}
\usepackage{xcolor}
\usepackage[normalem]{ulem}
\newcommand{\jetcaf}{{\fontfamily{qcr}\selectfont JeTCAF}}
\newcommand{\tcaf}{{\fontfamily{qcr}\selectfont TCAF}}

\newcommand{\tbabs}{{\fontfamily{qcr}\selectfont TBABS}}
\newcommand{\zpcfabs}{{\fontfamily{qcr}\selectfont ZPCFABS}}
\newcommand{\gabs}{{\fontfamily{qcr}\selectfont GABS}}
\newcommand{\gauss}{{\fontfamily{qcr}\selectfont GAUSS}}

\begin{document}

   \title{Variable mass accretion and failed wind explain changing look phenomena in NGC\,1365}


   \author{Santanu Mondal,
          \inst{1}
          Tek P. Adhikari,\inst{2} 
          Krzysztof Hryniewicz,\inst{3}
          C. S. Stalin,\inst{1}
          \and
          Ashwani Pandey\inst{1}
          }

   \institute{Indian Institute of Astrophysics, II Block, Koramangala, Bangalore 560034, India\\
            \email{santanuicsp@gmail.com}
         \and
             Inter-University Centre for Astronomy and Astrophysics, Pune, Maharashtra 411007, India\\
             \email{tek@iucaa.in}
             \and
             National Centre for Nuclear Research, ul. Pasteura 7, 02-093 Warsaw, Poland\\
             \email{krzysztof.hryniewicz@ncbj.gov.pl}
             }

 
  \abstract
   {Changing look active galactic nuclei (CLAGNs) show complex 
nature in their X-ray spectral shape and line of sight column density variation.  
The physical mechanisms responsible for these variations are unclear. 
Here, we study the spectral properties of a CLAGN, NGC\,1365  
using combined {\it XMM-Newton} and {\it NuSTAR} observations to understand the CL behavior. 
The model fitted mass accretion rate varied between $0.003\pm 0.001$ and $0.009\pm0.002$ $\dot M_{\rm Edd}$ and the dynamic corona changed from $28\pm 3$ to $10\pm1$ $r_g$.
We found that the variable absorption column density correlates with the mass accretion rate 
and the geometry of the corona. The derived wind velocity was sufficiently low compared to the escape velocity to drive the wind away from the disc for the epochs when column densities were high. This suggests that the high and variable absorption can be due to failed winds from the disc. Our estimated ratio of mass outflow to inflow rate from the inner region of the disc lies between $0.019\pm0.006$ and $0.12\pm0.04$. From spectral fitting of the combined data, we found the mass of the central black hole to be constant $4.38\pm0.34 - 4.51\pm0.29 \times10^{6} M_\odot$, consistent with earlier findings. The confidence contours of $N_H$ with other model parameters show that the model fitted parameters are robust and non-degenerate. Our study construed that the changing accretion rate, which is a fundamental physical quantity and the geometry of the corona driving the CL phenomena in NGC\,1365. The physical picture considered in this work connects both variable continuum and variable absorbing medium scenarios.}

\authorrunning{Mondal et al.}
  \maketitle

\section{Introduction}
An active galactic nucleus (AGN) is powered by the matter accreting onto the central 
supermassive black hole (SMBH) at the center of a galaxy. 
The radiation coming from the accreting AGN systems shows complex
features in their X-ray spectra. The components (e.g. hard, soft, or soft excess etc.) of the spectra come from different regions within the system. 
Some fraction of the hard radiation heats up the disc and generates emission lines, while some fraction gets absorbed
by different clouds at the line of sight (LOS). The variable absorption of the continuum coming from the central inner disc  can be due to the presence of variable clouds or due to the winds from the disc.

The X-ray absorbing column density changes 
dramatically by an order of magnitude over
a few years in different Seyfert galaxies \citep{Risalitietal2002}, which disfavors 
the homogeneous nature of the absorbers. 
Changing look (CL) AGN are those sources that show 
extreme variation in the column density of the X-ray absorber, with rapid transitions between Compton-thin (N$_H < 10^{24}$ cm$^{-2}$) and Compton-thick (N$_H > 10^{24}$ cm$^{-2}$) regimes \citep[e.g.][]{Riccietal2016}. 
CLAGN though first discovered serendipitously \citep{LaMassaetal2015} are now considered as a separate class of AGN that transit from type 1 to 2 and vice versa within a timescale of decades or years \citep{Veronetal1980,Maiolino1995,Denneyetal2014,Yangetal2018,Kimetal2018}. In \citet{Riccietal2020}, the CL phenomena was shown for 1ES 1927+654 can be due to rapid changes in the innermost regions of accreting SMBHs.

According to models, the timescale of CL phenomenon within months or years can be due to the variation in absorbing material that appears and disappears along the LOS, this supports the quick variation in column density \citep{Risalitietal2009}. On the contrary, many CLAGNs showed little variation in column density \citep{McElroyetal2016,NodaDone2018}, therefore CL nature could also be due to 
other physical processes within the system. The weak or low optical polarization 
of CLAGNs suggests that the transition from type 1 to type 2 can not be attributed to the putative dust obscuration \citep[e.g.][]{Hutsemekers2019}. The mid-infrared and optical analysis 
of \citet{Shengetal2017} do not favor the obscuration scenario and support that the 
CL may be caused by the intrinsic variation of accretion process. The work of \citet{NodaDone2018}
suggests CL as accretion process originated phenomenon where radiation pressure should be 
more important compared to gas pressure. The radiation pressure driven instability is also 
a possibility to explain the observed timescale of CL phenomenon \citep{Sniegowska2019}. 

NGC\,1365 is a Seyfert 1.8 \citep{MaiolinoReike1995} galaxy at z=0.0055 with a $10^{6.5} M_\odot$ BH \citep{Risalitietal2009}. \citet{CombesEtal2019} estimated the mass of $3.98 \times 10^6 M_\odot$ from the molecular gas dynamics using ALMA data.  This spiral galaxy is intensely studied by \citet[][for a review]{Lindbladreview1999}.
The source is a CLAGN due to its Compton-thin to Compton-thick transition, that can be due to variation in the LOS absorber rather than to extreme intrinsic emission variability \citep{MattEtal2003,Risalitietal2005}.
The spectral components of NGC\,1365 in X-ray consists of several distinct features e.g. Compton hump $> 10$ keV, soft excess at low energy (below 3 keV), strong absorption features, relativistic iron and emission lines. The source also has bi-conical outflow from the nucleus \citep{Edmundsetal1988}. \citet{Sandqvistetal1995} suggested that the AGN has a weak radio jet in this source. A [O III] line is seen in the outflow \citep{Storchi-Bergmann1991,Veilleuxetal2003,SharpBlandHawthorn2010}. The emission lines observed for this source are dominated by collisionally ionized gas \citep{Guainazzietal2009}, however, it can also be hybrid \citep[photoionised+collisionally ionised;][used {\it XMM-Newton} and {\it Chandra} data]{Whewelletal2016}. NGC\,1365 has some important characteristics, when it is observed in the Compton-thin state, its X-ray spectrum  shows the presence of strong absorption lines due to Fe xxv(both the He-$\alpha$ and He-$\beta$) and Fexxvi(both the Ly$\alpha$ and Ly$\beta$ components); these lines are blue shifted by v$\sim$3000\,km s$^{-1}$ \citep{Risalitietal2005,BrennemanEtal2013}. These features can be originated due to the presence of highly ionised outflowing wind and possibly originated from the accretion disc. All these features are occasionally observed in Seyfert galaxies, however, it is unlikely to see all of them in a single source. Later, \citet{ConnolyEtal2014} examined that the long-term variability of the X-ray absorber could be originated from the changes in mass accretion rate of NGC\,1365. The analysis of {\it Suzaku} observation of the source suggested the presence of a partial covering and ionized absorber using photoionisation models \citep{GoffordEtal2013,GoffordEtal2015}. \citet{BraitoEtal2014} evidenced that two ionised absorbers were needed to fit the RGS spectra of {\it XMM-Newton} and detected a possible P-Cygni profile of the Mg\,XII Ly$\alpha$ line which is associated with mildly ionized absorber indicative of a wide angle outflowing wind. At the same time, using the {\it Chandra} data, \citet{NardiniEtal2015} estimated the location of the photoionised gas based on some evidence of broadening of the emission lines.
The central active nucleus is surrounded by a molecular torus with both narrow 
and broad emission line components in the emitted spectrum and relativistic iron line \citep{Risalitietal2009}. 
The variable absorbing medium along the LOS was observed by \citep{Riversetal2015} using combined {\it XMM-Newton} and {\it NuSTAR} data. 
The relativistic line feature estimates the maximal spin of the BH $> 0.97$ \citep{Waltonetal2014}. 
\citet{Karaetal2015} estimated soft lag due to the presence of eclipsing cloud along the line of sight using {\it NuSTAR} observations. The star formation and gas accretion to the central object were studied using near-infrared (NIR) data by \citet{Fazelietal2019} and \citet{Gaoetal2021} to check for more recent star formation activities and outflow structure of the source.

Despite the source has been studied extensively in different energy bands, the understanding of CL phenomena is still lacking due to shortcoming of physical picture in earlier studies. Most of them phenomenologically explained that the variable column density is due to the appearance/disappearance of some clouds along the LOS and  none of the previous studies self-consistently considered the accretion-ejection scenario to take into account for the CL phenomena. Therefore, in this work, we aim to address that phenomena after estimating fundamental physical quantities of accretion e.g. mass accretion rate, corona geometry etc., considering the beforehand observational characteristics such as Compton hump, jets, winds, and variable continuum in NGC\,1365 system. For that, we used joint {\it XMM-Newton} and {\it NuSTAR} data during 2012 and 2013. 
This paper is organized as follows: in Section 2 we explain the observations and data reduction
procedures, in Section 3, we discuss the modelling, various analysis, and data fitting carried out using different models. A brief conclusion is given in the final Section.

\section {Observation and Data analysis}\label{sec:Observation}

We used joint {\it XMM-Newton} and {\it NuSTAR} \citep{Harrisonetal2013} observations of NGC\,1365 during 2012 and 2013. The log of observations is given in \autoref{table:observation}.

\begin{table}
\scriptsize
\centering
\caption{\label{table:observation} Log of observations. Here, X=6000204600, and Y=0692840 are the initials of the observation IDs.}
\begin{tabular}{ccccccc}
\hline
  {\it NuSTAR} &Exp. & {\it XMM-Newton} &Exp. &Epoch   & Date   & MJD\\
            &    (ks)   &          &(ks)      &    &   &   \\
\hline
X3&41 &Y201&118 &A  &2012-07-26 &56134 \\
X5&66 &Y301&108 &B  &2012-12-24 &56285 \\
X7&74 &Y401&93  &C  &2013-01-23 &56315 \\
X9&70 &Y501&116 &D  &2013-02-12 &56335 \\
\hline
\end{tabular}
\end{table}

To analyze the {\it XMM-Newton} data, we used the Science Analysis System (SAS) version 19.1.0 and followed the standard procedures given in the SAS data analysis threads\footnote{\url{https://www.cosmos.esa.int/web/xmm-newton/sas-threads}}. We first reprocessed the Observation Data Files (ODF) using the {\sc epproc} routine to generate the calibrated and concatenated event lists. We filtered the concatenated event lists to remove the periods of high background activity. We then extracted the source and background spectra using the {\sc evselect} routine. We used a circular region of radius $32^{\prime\prime}$ for both the source and the background. The background region was in the same CCD but away from the contamination of the point sources. We used {\sc rmfgen} and {\sc arfgen} tasks to generate the redistribution matrices (.rmf)  and the ancillary (.arf) files. Finally, we rebinned the spectra using the task {\sc specgroup} to have at least 25 counts in each bin.

The {\it NuSTAR} data were extracted using the standard 
{\sc NUSTARDAS v1.3.1}
\footnote{\url{https://heasarc.gsfc.nasa.gov/docs/nustar/analysis/}} software. We 
ran {\sc nupipeline} task to produce cleaned event lists and 
{\sc nuproducts} to generate the spectra. We used a region of 
$80^{\prime\prime}$  for the source and $100^{\prime\prime}$ for the background 
using {\sc ds9}. The data were grouped by {\it grppha} command, with a 
minimum of 30 count in each bin. For the analysis of each epoch of 
observation, we used the data of joint {\it XMM-Newton} and {\it NuSTAR} in the energy range of 0.5$-$75 keV.

We used {\sc XSPEC}\footnote{\url{https://heasarc.gsfc.nasa.gov/xanadu/xspec/}} \citep{Arnaud1996} version 12.11.0 for spectral analysis. 
Each epoch of observation was fitted using an accretion disc 
based two component advective flow (\tcaf~\citet{ChakrabartiTitarchuk1995}) model including jet (\jetcaf~\citet{MondalChakrabarti2021}). The present model takes into account the radiation processes at the base of the jet, discussed later in detail. We used the absorption model 
\tbabs \citep{Wilmsetal2000} for the Galactic absorption, keeping the hydrogen column density fixed to $1.3\times 10^{22}$cm$^{-2}$ \citep{Kalberlaetal2005} during the fitting. We used chi-square statistics for the goodness of the fitting. We employed source redshift here and left $N_H$ as a free parameter to estimate the absorbing column density responsible for the overall spectral distortion.

\section{Modelling}\label{sec:Modelling}

\begin{figure}
\centering{
\hspace{-1.2cm}
\includegraphics[height=6.0truecm]{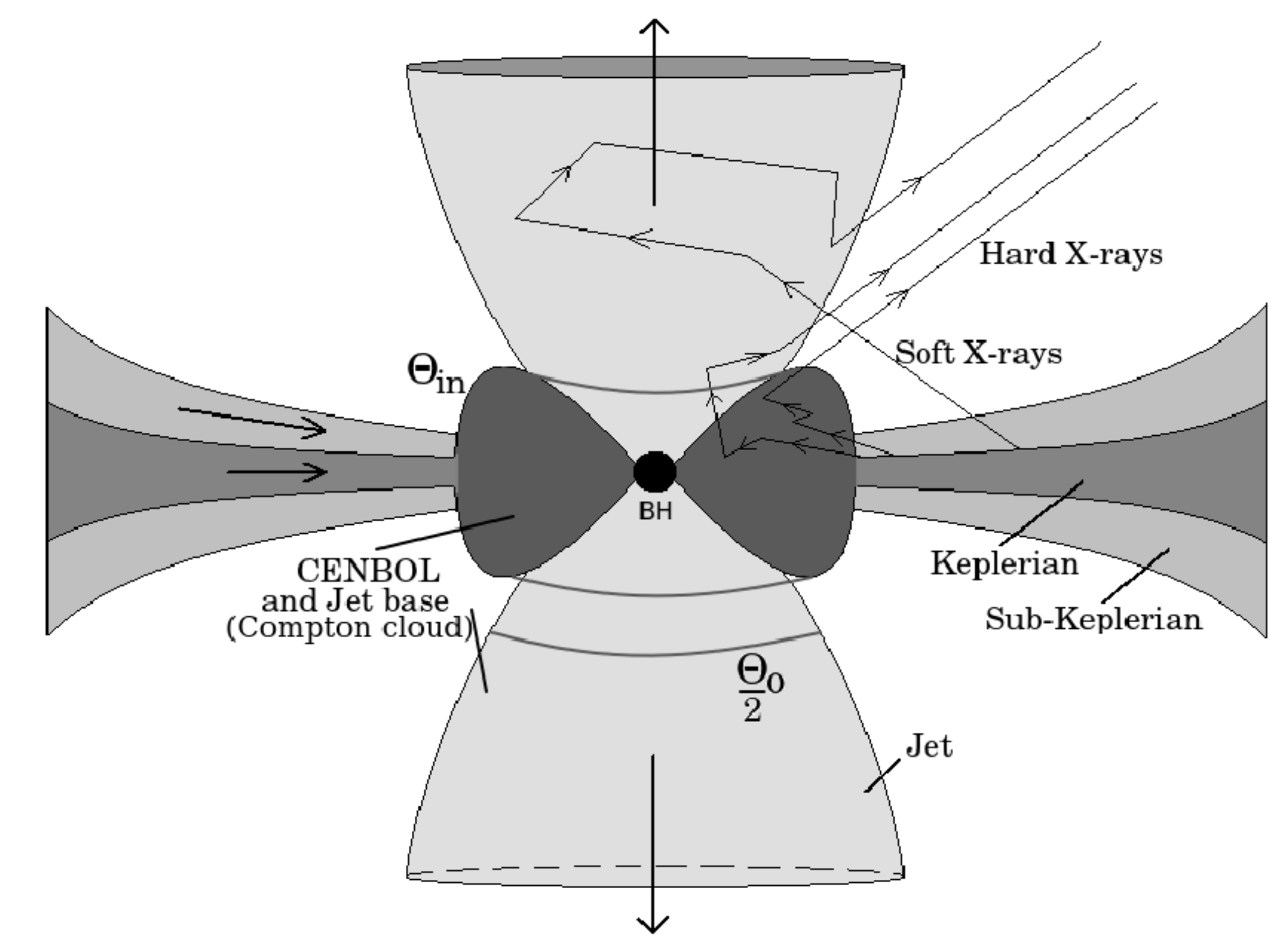}}
\caption{Illustration of the \jetcaf~model. The zig-zag arrows show the scattering of photons from the disc and jet. The $\Theta_o$ and $\Theta_{\rm in}$ are the solid angle subtended by the outflow and inflow. The figure is adopted from \citet{MondalChakrabarti2021}.
} 
\label{fig:modelCartoon}
\end{figure}

For the spectral fitting, we used the accretion-ejection based \jetcaf~model \citep{MondalChakrabarti2021} along with
multiplicative models e.g. partial covering fraction absorption (\zpcfabs), and Gaussian absorption line (\gabs). The \jetcaf\, model geometry and flow configuration is illustrated in \autoref{fig:modelCartoon}. 
In this model, the accretion disc has two components, one is a Keplerian disc sitting at the equatorial plane which is the source of soft photons, and a sub-Keplerian 
halo (hot component) sitting above and below the Keplerian disc. The second component forms the shock in accretion disc \citep{Chakrabarti1989} after satisfying Rankine-Hugoniot conditions.
This shocked region behaves as the centrifugal pressure supported boundary layer (CENBOL) of the corona. The same region is also producing thermally driven jets. Matter falling onto the black hole, piles up inside the shocked (post-shock) region. The \jetcaf\, model takes into account the radiation mechanisms at the base of the jet and the bulk motion effect by the outflowing jet on the emitted spectra. This model requires six parameters (if the mass is considered as a free parameter) which are (i) mass of the BH ($M_{\rm BH}$), (ii) disc accretion rate 
($\dot m_{\rm d}$), (iii) halo accretion rate ($\dot m_{\rm h}$), (iv) size of the dynamic corona or the location of the shock ($X_{\rm s}$ in $r_g=2GM_{\rm BH}/c^2$ unit), (v) shock compression ratio ($R$), and (vi) ratio of the solid angle subtended by the outflow to the inflow ($f_{\rm col}=\Theta_o/\Theta_{\rm in}$). 

The emergent spectrum from \jetcaf\,has four components, (1) multicolour blackbody spectrum which is coming from the disc, (2) the hard radiation; from the upscattering of the soft photons from the disc by the hot corona, (3) the second hard component at the shoulder of the black body bump is due to scattering of soft photons from the disc by the base of the jet, (4) downscattering of hard radiation from the corona by the bulk motion of the jet produces excess above 10\,keV \citep{TitarchukShrader2005,MondalChakrabarti2021}. This component also fits the Compton hump that are observed in reflection models. The current \jetcaf\, model is an updated version (including jet) of \tcaf\, model, which has been successfully applied to low mass black hole X-ray binaries \citep[][and references therein]{Debnathetal2014,Mondaletal2014,MondalChakrabarti2019} and to AGNs  \citep[e.g.][]{Nandietal2019,MondalStalin2021} to fit their spectra and to infer the underlying accretion dynamics. In this model the same corona, which is producing hard radiation, also launching jet at its base. The jet/mass outflow rate is coming after solving a series of coupled hydrodynamic equations, thus, it naturally connects disc and jet. As NGC\,1365 evidenced the presence of both accretion discs, jets/outflows and Compton hump $> 10$keV, therefore, it is worth applying \jetcaf\, model to study the CL phenomena. For the first time, we are applying \jetcaf\, model as a physical model to understand the CL phenomena, and we believe it has the potential to explore the observed accretion-ejection features in NGC\,1365. We note that, the present model does not include detailed line emission/absorption properties.

The spectrum of NGC\,1365 showed complex features including distortions between 3-7 keV, which can be due to varying absorber covering the source, that substantially changes the shape of the spectrum below 10\, keV \citep{Risalitietal2009}. It also showed broadening of Fe line $\sim$6\,keV in combined {\it XMM-Newton} and {\it NuSTAR} data due to relativistic effects near the inner region of the disc \citep{Risalitietal2013}. Below $\sim 3$\,keV energy, the {\it XMM-Newton} data showed diffuse thermal contribution and significant excess below $\sim$2\,keV, the origin of which is not clear yet. However, it can be originated by the up-scattering of seed photons from the disc by some warm, optically thick medium \citep[see for recent study][and references therein]{MehdipourEtal2021}. As the current \jetcaf\,model does not include relativistic effects, therefore for the skewing and broadening of the Fe line shapes near the innermost region of the disc, we used \gauss~model to take into account those effects. Furthermore, we focus on the accretion parameters and the geometrical configuration of the corona that generate variable emission, which can be successfully obtained by analyzing the combined {\it XMM-Newton} and {\it NuSTAR} data in the range of 0.4-75\,keV. In this range all spectra showed both emission and absorption features, those are taken into account by \gauss~and \gabs~models. One Gaussian component is used at $\sim 6.4$\,keV for FeK line and the second one is used at $\sim 0.8$\,keV to fit the broad line feature at low energy. The other two/three Gaussian components are applied at different energy regions throughout the spectra, however, all are used below 6\,keV. The origin of (Soft) Gaussian components below 2\,keV is not clear yet, however, it can be originated from distant reflection \citep{MislavEtal2018}.  \zpcfabs\,model is used for addressing the source intrinsic absorption, partial covering and to estimate the column density along the LOS. 
The complete model reads in  {\it XSPEC} as  
\\
\tbabs*\zpcfabs*\gabs[2]*(\gauss[4-5]+\jetcaf).

The \zpcfabs\, model is not a good representation for an ionizing medium, as it does not include the effects of ionization, to generate the emission/absorption lines. For the ionization rate estimation, freely available photoionization software (e.g. {\sc cloudy} or {\sc xstar} etc.) can be appropriate, as extensively discussed in the literature \citep[see][and references therein]{GoffordEtal2013,BraitoEtal2014,GoffordEtal2015}. However, the current paper aims to quantify the accretion-ejection parameters assuming that accretion rate and corona properties might be triggering CL phenomena, therefore the detailed ionization modelling is beyond the scope of the paper.

\section{Results and Discussions} \label{sec:Results}
\begin{figure*}
\centering{ 
\includegraphics[height=9truecm,angle=270]{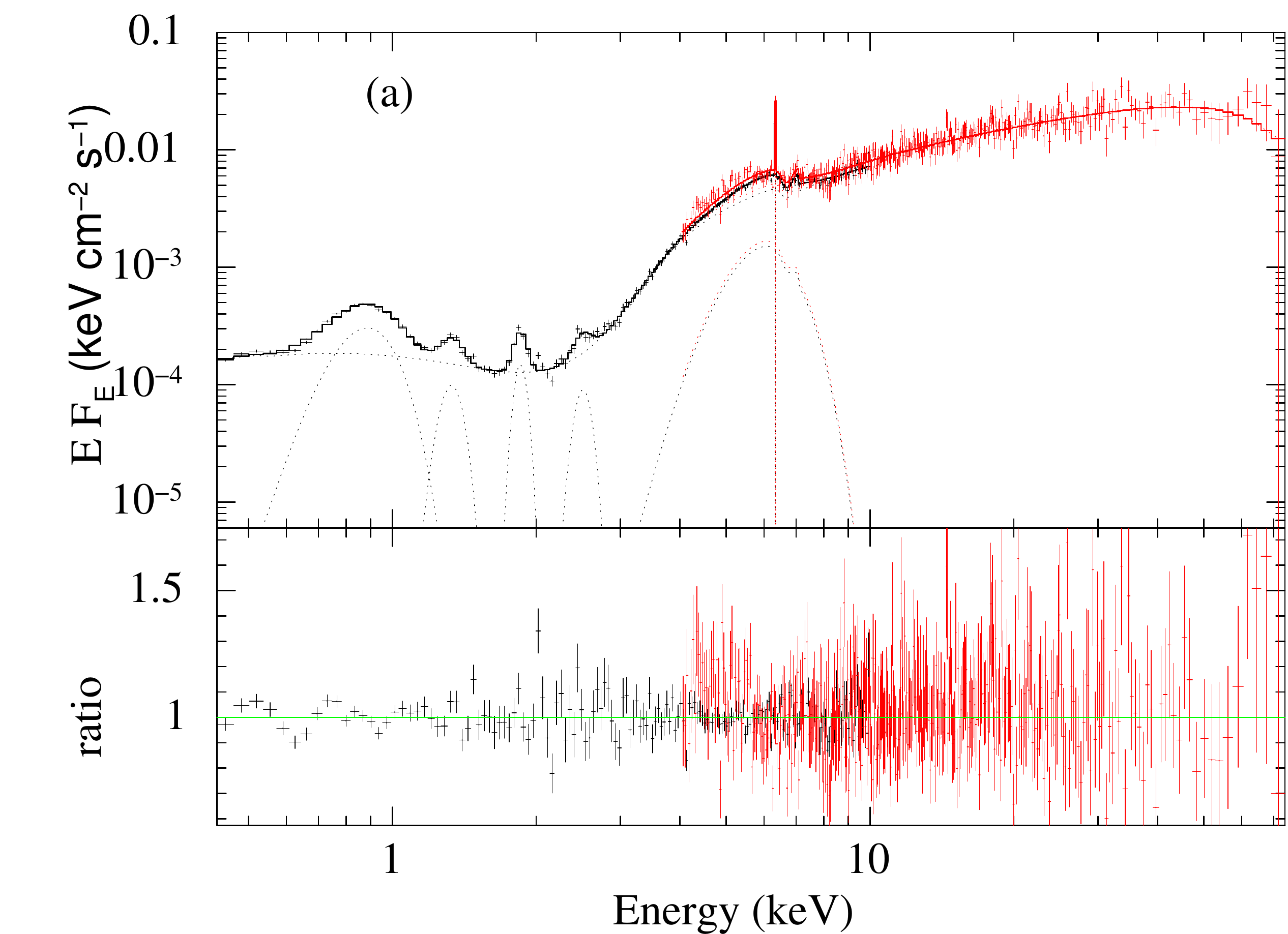}
\hspace{-0.5cm}
\includegraphics[height=9truecm,angle=270]{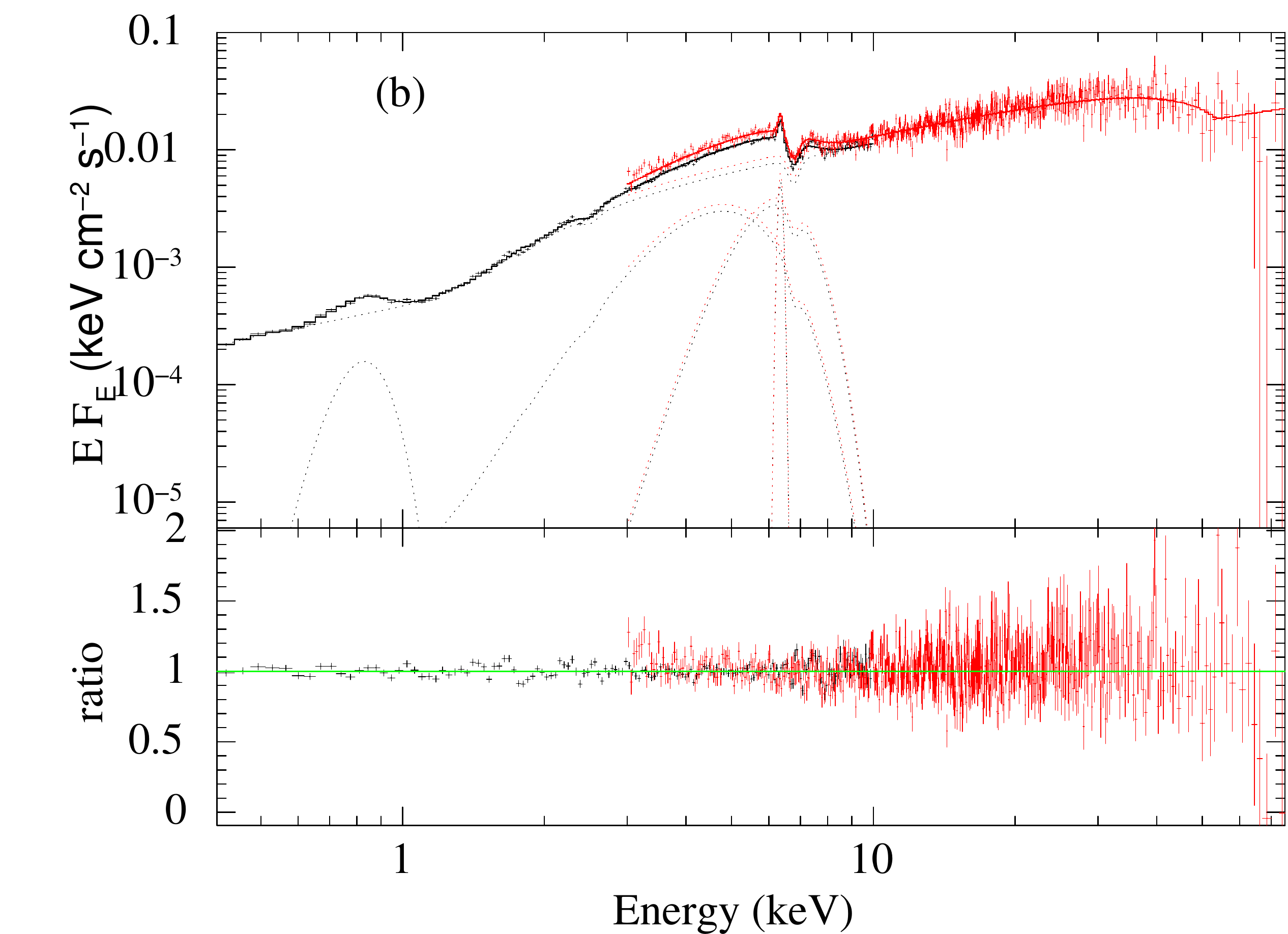}} \\
\centering{ 
\includegraphics[height=9truecm,angle=270]{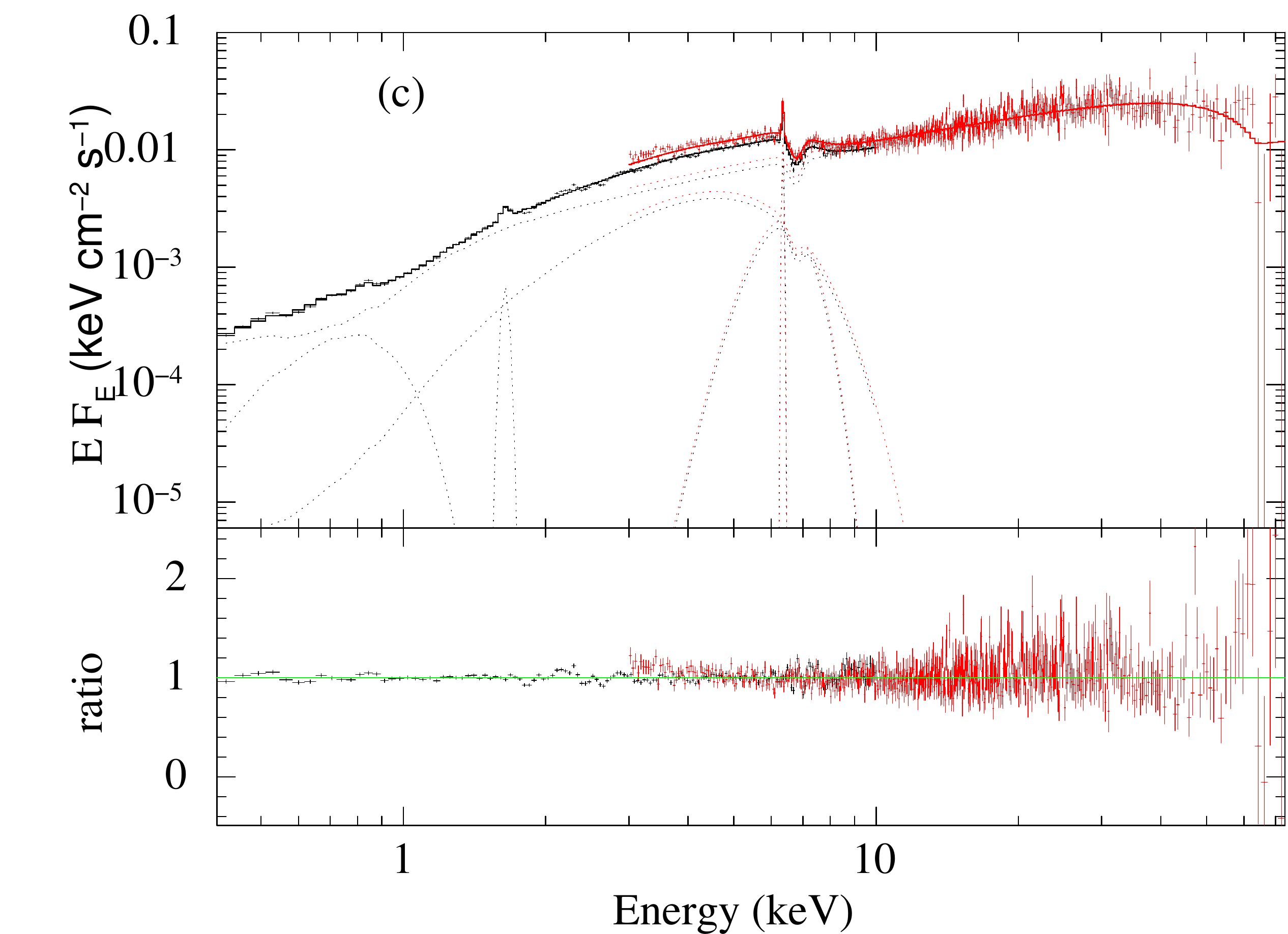}
\hspace{-0.5cm}
\includegraphics[height=9truecm,angle=270]{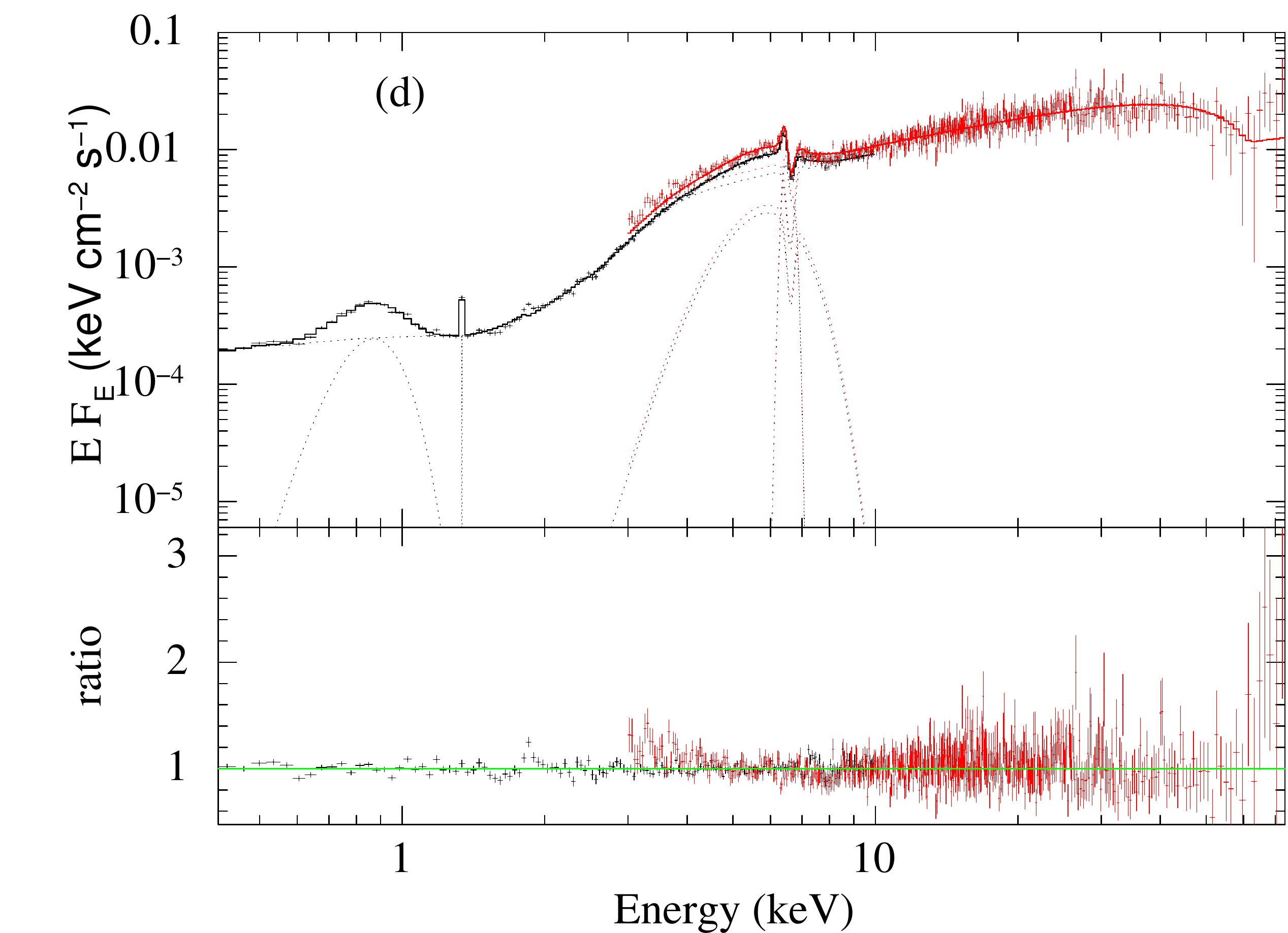}}
\caption {\jetcaf~model fitted 0.4$-$75.0 keV spectra for all four epochs of NGC\,1365 during {\it XMM-Newton} and {\it NuSTAR} era. Here, (a-d) corresponds to
epochs A to D. The different Gaussian lines are used for emission/excess at different energy range including at $\sim$6.4 keV. The model fits show the absorption line and spectral shape change in each epoch, along with a PCygni profile in the FeK band.}
\label{fig:SpecFits}
\end{figure*}

\begin{figure}
\centering{
\includegraphics[height=9.0truecm,trim={0.4cm, 0.9cm, 1.5cm, 2.0cm}, clip]{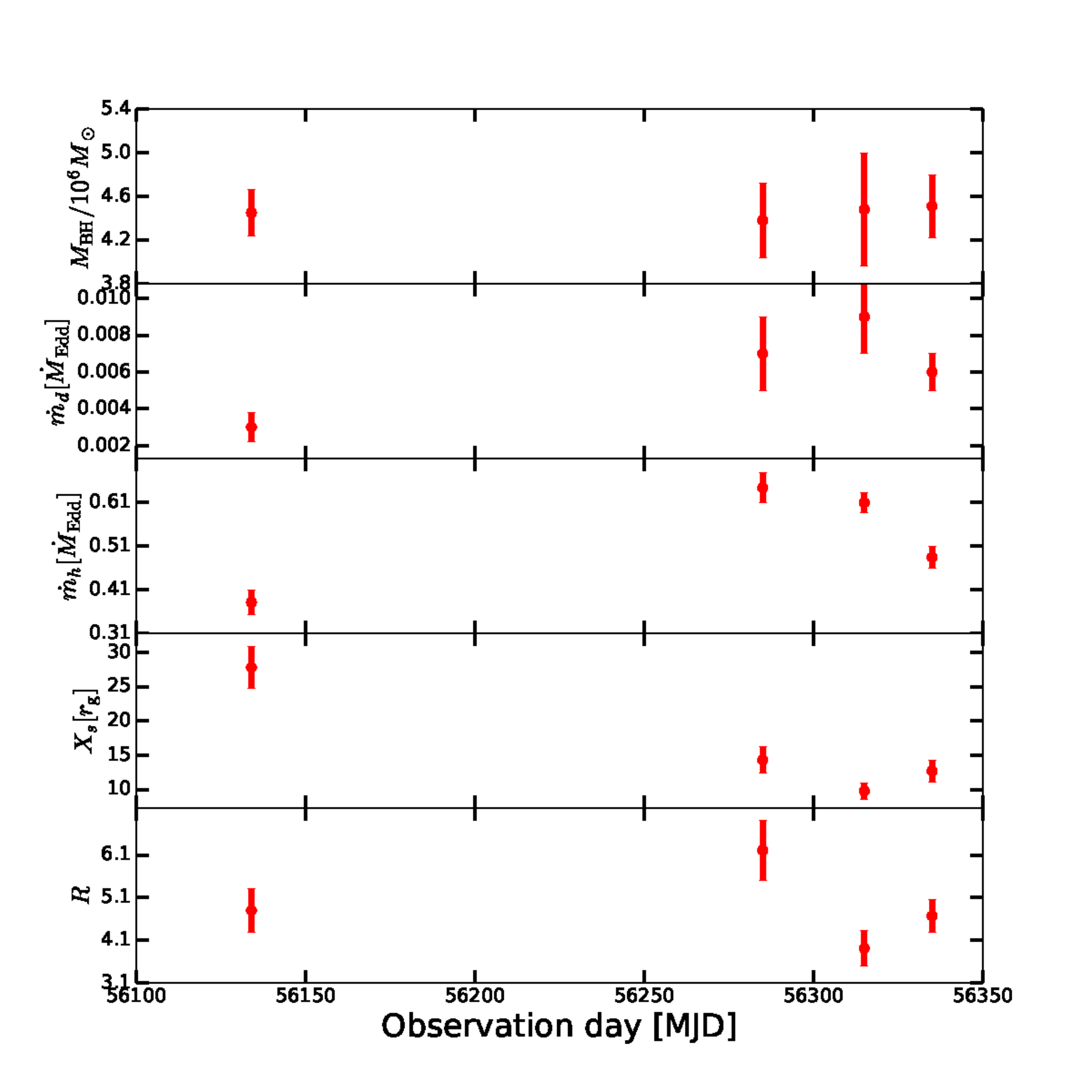}}
\caption{Variation of \jetcaf~model fitted parameters with observational epochs.
} 
\label{fig:FittedPars}
\end{figure}

\autoref{fig:SpecFits} shows the spectral fitting to the data for four epochs. In all the spectra, we found complex features including absorption lines, skewing of FeK line, soft excess, and some excess at high energies, which can be due to the bulk motion of the outflowing
material from the central region or can be due to reflection \citep[as reported by][and references therein]{Riversetal2015}. The model fitted parameters are given in \autoref{table:JetcafResults}. The variation of the \jetcaf\,model fitted parameters are shown in \autoref{fig:FittedPars}. The top panel shows the variation of mass which is constant during the observation period, as expected, however, other parameters changed significantly from second top to bottom panel. The disc or Keplerian accretion rate was $\leq$ 1\% of the Eddington rate, however, changed significantly between epochs A to D (second panel). \citet{BraitoEtal2014} estimated the mass accretion rate for this source, which is a factor few higher than our estimates. It can be due to either strong dependence of their estimates on the ionized winds and the covering factor they used in their model or due to using higher mass of the BH ($10^7 M_\odot$) and a different accretion efficiency (0.06).
The low-angular momentum, hot flow rate was always higher than the Keplerian one. We also noticed that the size of the dynamic corona shrunk from 28 to 10 $r_g$. As the size of the corona was changed significantly, and the disc accretion rate
was low compared to the halo accretion rate, these are the indication of hard/intermediate spectral state, if the spectra of AGN allow a similar classification as in low mass black hole binaries. 
Our model fitted shock compression ratio ($R$) varies between 3.9 to 6.2. As the presence of jet was confirmed from [{\sc OIII}] 
line that the source has bi-conical outflows \citep{Edmundsetal1988}, we therefore estimated the mass outflow/inflow rate (dimensionless quantity) using  the relation, $R_{\dot m}=\frac{\dot M_{\rm out}}{\dot M_{\rm tot}}=f_{\rm col}f_0^{3/2}\frac{R}{4} exp\left[\frac{3}{2} -f_0\right]$,
where, $f_{\rm 0}=\frac{R^2}{R-1}$ \citep{Chakrabarti1999}, $\dot M_{\rm tot}=(\dot m_d+R\times \dot m_h)$ \citep{MondalEtal2014ApSS}, and $\dot M_{\rm out}$ is the mass outflow rate in $M_\odot yr^{-1}$ respectively. The $f_{\rm col}$ value varied between 0.22 to 0.4 (column 7 in \autoref{table:JetcafResults}. Our calculated values lie in the range between $0.019\pm0.006$ to $0.12\pm0.04$. This gives the mass outflow rate $\sim 0.047$ M$_\odot$ yr$^{-1}$,  which is in agreement ($\sim 0.067$ M$_\odot$ yr$^{-1}$) to the independent optical and {\it Chandra} imaging studies\citep{Venturietal2018} and the {\it XMM-Newton}-RGS data analysis \citep{BraitoEtal2014}.  

Treating BH mass as a free parameter in \jetcaf~we found BH mass is constant (see \autoref{fig:FittedPars}) during the combined {\it XMM} and {\it NuSTAR} data fitting, that comes out to be $4.38\pm0.34 - 4.51\pm0.29 \times10^{6} M_\odot$, closely matches ($4.47 \times 10^6 M_\odot$) with the estimation by \cite{Onorietal2017} and from the molecular gas dynamics ($3.98 \times 10^6 M_\odot$) using ALMA data by \citet{CombesEtal2019}.

\begin{figure}
\centering{
\includegraphics[height=4.5truecm,trim={2.0cm, 0.0cm, 0.0cm, 1.5cm}, clip]{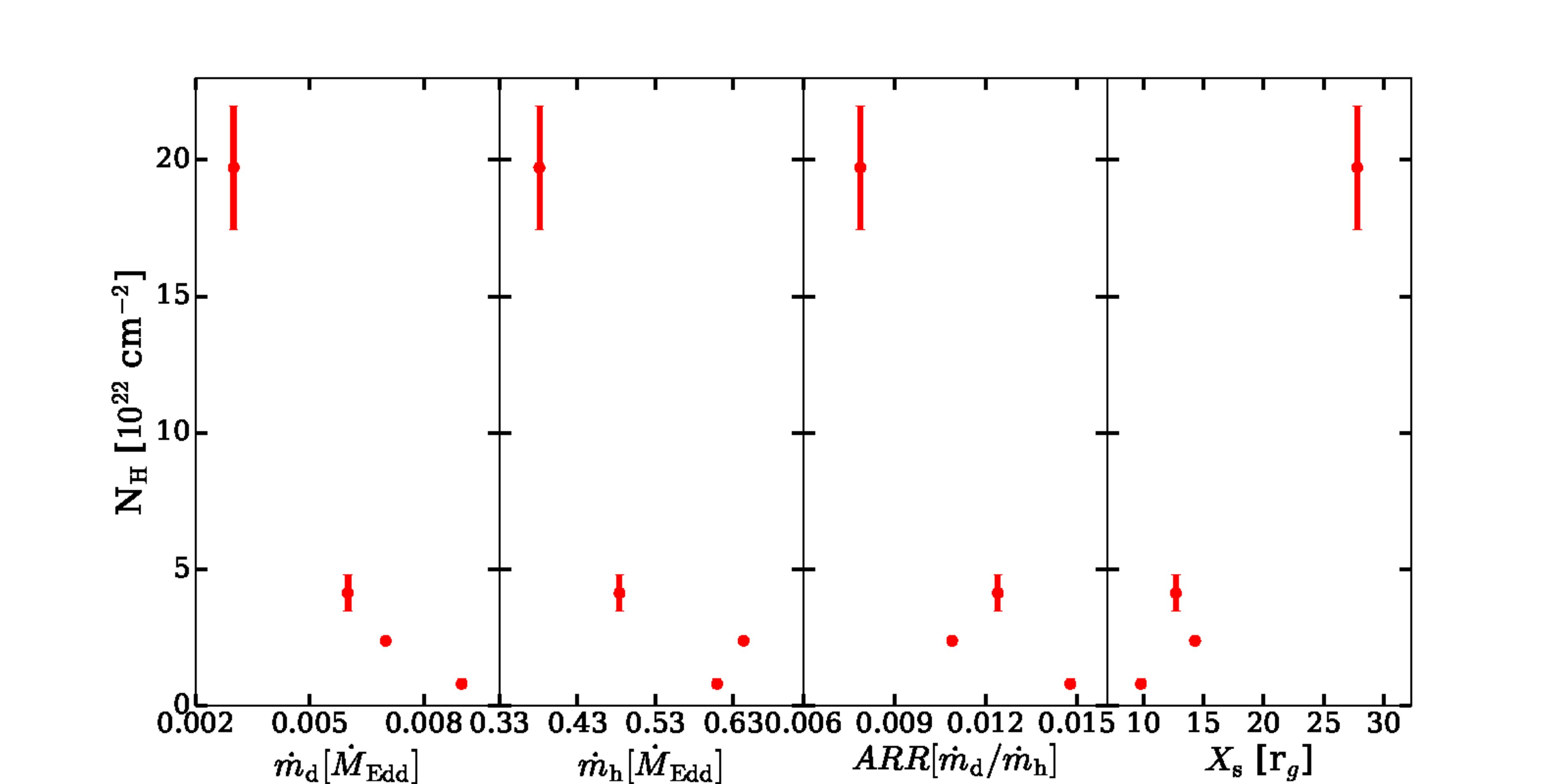}}
\caption{$N_{\rm H}$ decreases with increasing accretion rates, and an opposite profile is observed with the geometry of the corona.
} 
\label{fig:correlation}
\end{figure}

In addition to accretion disc parameters, the \zpcfabs~model fitted $N_{\rm H}$ varied (in column 8 of \autoref{table:JetcafResults}) by more than an order of magnitude between epochs A to C, and the partial covering fraction ($c_f$) was relatively high (see column 9) in all four epochs, however, constant within the error bar.
The variation of $N_{\rm H}$ with \jetcaf\,model parameters are shown in \autoref{fig:correlation}. The left panel shows its variation with $\dot m_{\rm d}$. A similar variation is observed with the halo rate and accretion rate ration (ARR=disc rate/halo rate). This relative ratio of accretion rates is an indicator of the ratio of disc and corona flux contribution. As the ratio is always $<1$, implies dominant contribution of hot flow rate; thus the radiation coming from the corona is hard. It is quite certain that changing the geometry and accretion flow properties can change the intensity of the radiation emitting from the corona, since it reprocesses photons from the underlying accretion flow. This radiation when falling onto the disc can also influence the disc mass outflow, thus the column density, which can be seen from other panels in \autoref{fig:correlation}. The confidence contours of $N_H$ with \jetcaf\,model parameters are shown in \autoref{fig:ConfCont} in \autoref{sec:Contour}, which show that the parameters are non-degenerate and robust.

\begin{table*}
\centering
\caption{\label{table:JetcafResults} Best fitted model parameters are provided in this table. Here, N$_H$ and c$_f$ are hydrogen column density and dimensionless covering fraction. 
$E_a$ and $\sigma_a$ are the absorption line energy and width. The superscripts 1 and 2 correspond to the absorption line 1 and 2. }
\resizebox{\textwidth}{!}{\begin{tabular}{cccccccccccccc}
\hline
Epoch       &$M_{\rm BH}$ &$\dot m_{\rm d}$ & $\dot m_{\rm h}$ & $X_{\rm s}$ & R &$f_{\rm col}$&$N_H$&c$_f$&$E_a^1$&$\sigma_a^1$&$E_a^2$&$\sigma_a^2$&$\chi_{\rm r}^2/dof$ \\
	        & $\times 10^6 M_\odot$&$\dot M_{\rm Edd}$&$\dot M_{\rm Edd}$&$r_{\rm g}$& & &$\times 10^{22}$ cm$^{-2}$& &keV &keV &keV &keV & \\
\hline
A &$4.45\pm0.21$&$0.003\pm0.001$&$0.381\pm0.028$&$27.8\pm3.1$&$4.80\pm0.52$&$0.39\pm0.08$   &$19.72\pm2.27$&$0.79\pm0.17$ &$2.04\pm0.12$&$0.91\pm0.13$&$6.71\pm0.08$&$0.17\pm0.06$&641/506 \\
B &$4.38\pm0.34$&$0.007\pm0.001$&$0.643\pm0.034$&$14.3\pm1.9$&$6.21\pm0.51$&$0.22\pm0.03$     &$2.38\pm0.15$ &$0.77\pm0.03$ &$2.50\pm0.02$&$0.11\pm0.03$&$6.78\pm0.02$&$0.19\pm0.02$&1101/704 \\
C &$4.48\pm0.52$&$0.009\pm0.002$&$0.609\pm0.022$&$9.8\pm1.2$&$3.91\pm0.42$&$0.38\pm0.11$      &$0.80\pm0.18$ &$0.89\pm0.08$ &$1.87\pm0.06$&$1.74\pm0.07$&$6.76\pm0.02$&$0.21\pm0.03$&1214/707 \\
D &$4.51\pm0.29$&$0.006\pm0.001$&$0.484\pm0.024$&$12.7\pm1.6$&$4.67\pm0.38$&$0.40\pm0.09$     &$4.13\pm0.67$ &$0.87\pm0.04$ &$2.26\pm0.15$&$0.75\pm0.10$&$6.62\pm0.15$&$0.12\pm0.19$&959/688 \\
\hline
\end{tabular} }
\end{table*}

\subsection{Robustness of model parameters} \label{sec:Contour}
\begin{figure*}[h!]
\centering{ 
\includegraphics[height=6.5truecm,angle=270]{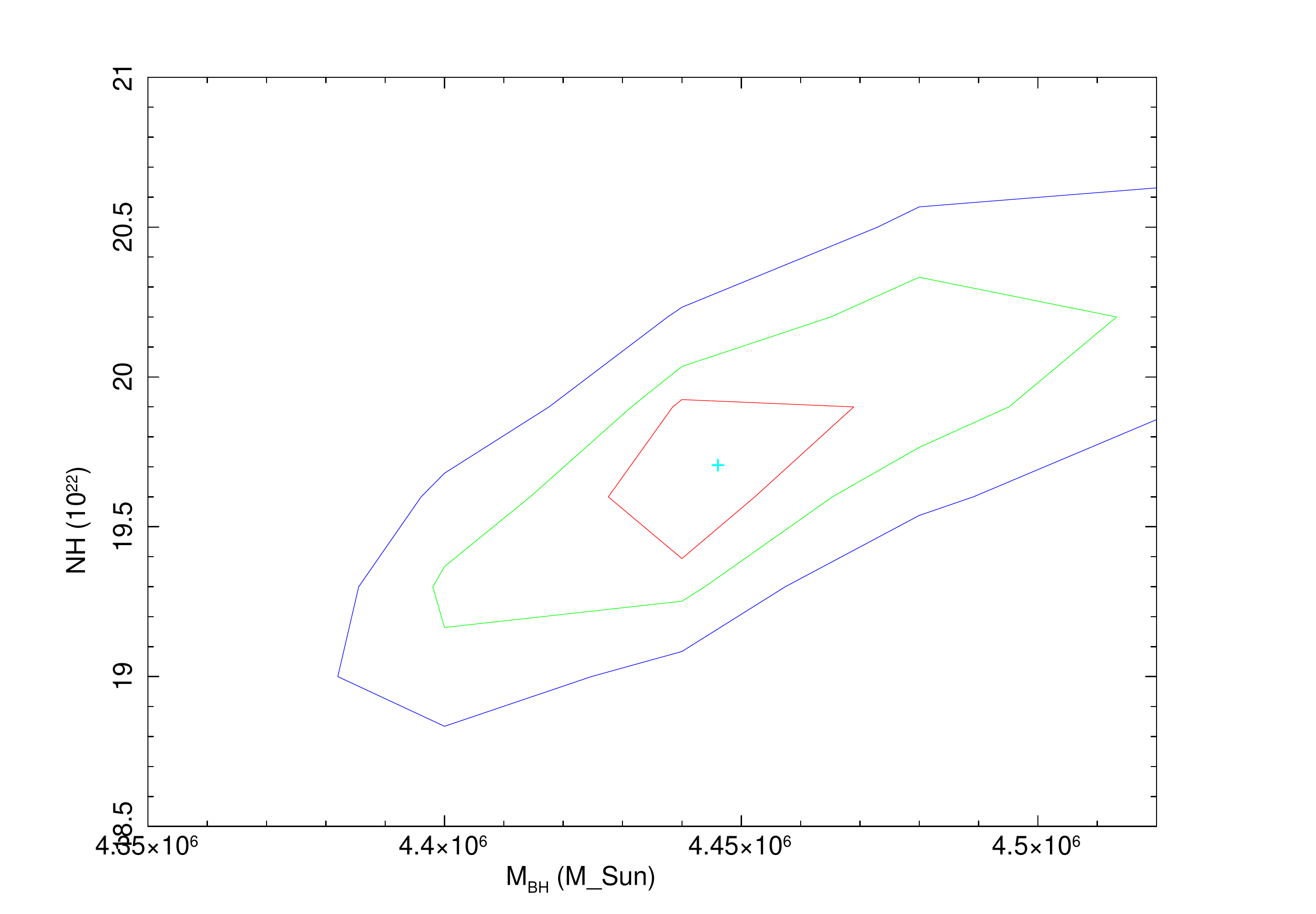}
\hspace{-1.0cm}
\includegraphics[height=6.5truecm,angle=270]{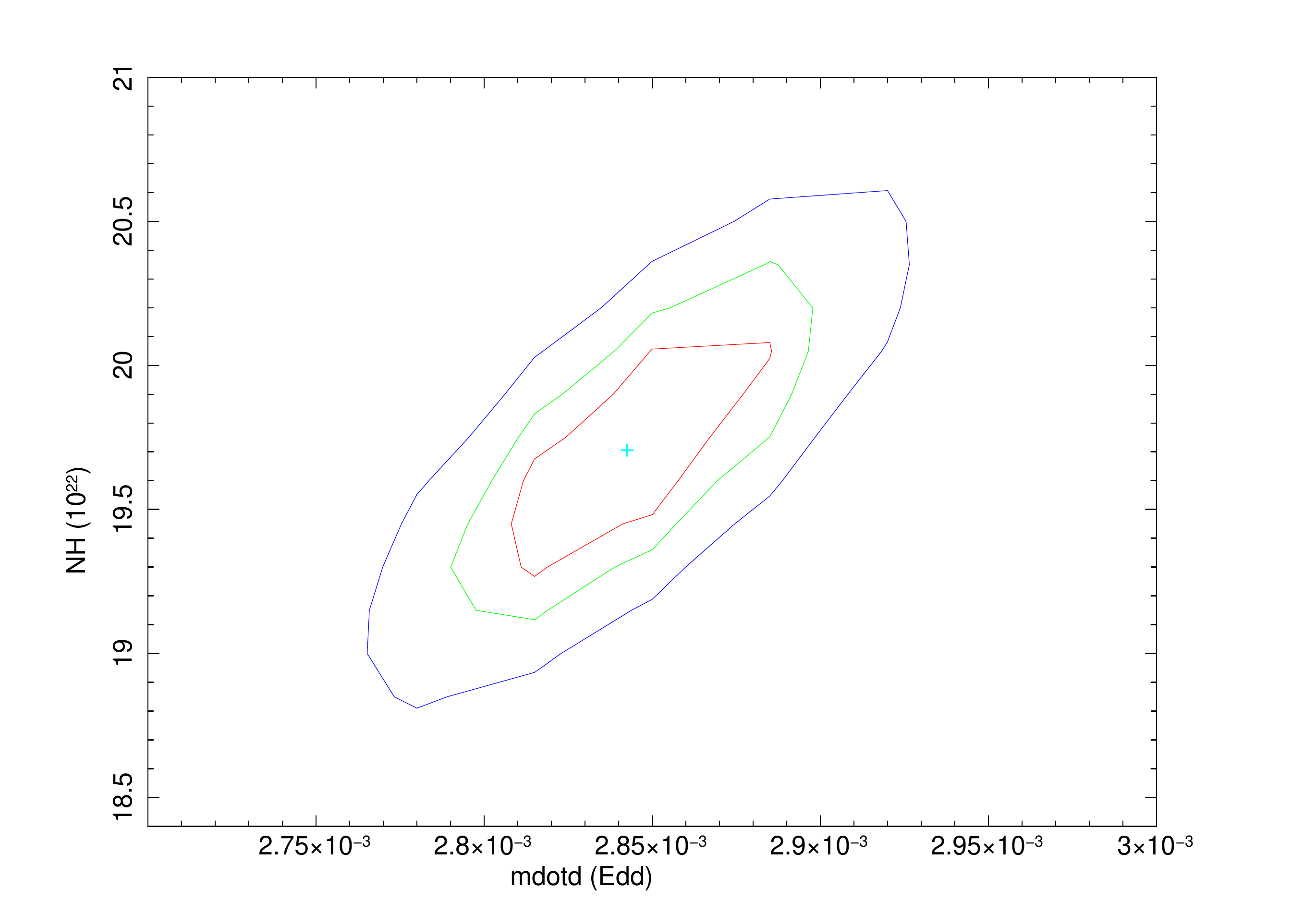}
\hspace{-1.0cm}
\includegraphics[height=6.5truecm,angle=270]{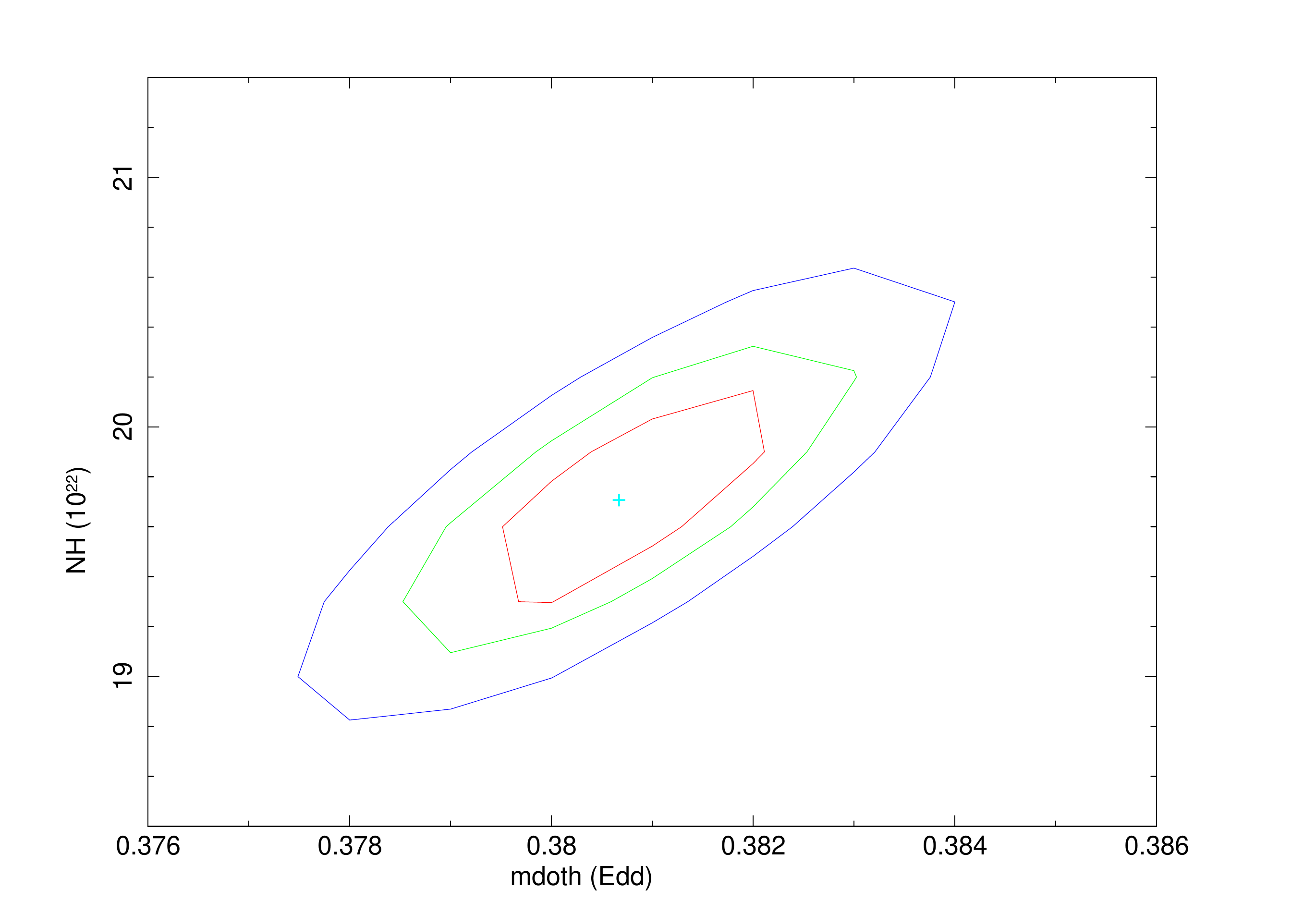}}\\
\centering{ 
\includegraphics[height=6.5truecm,angle=270]{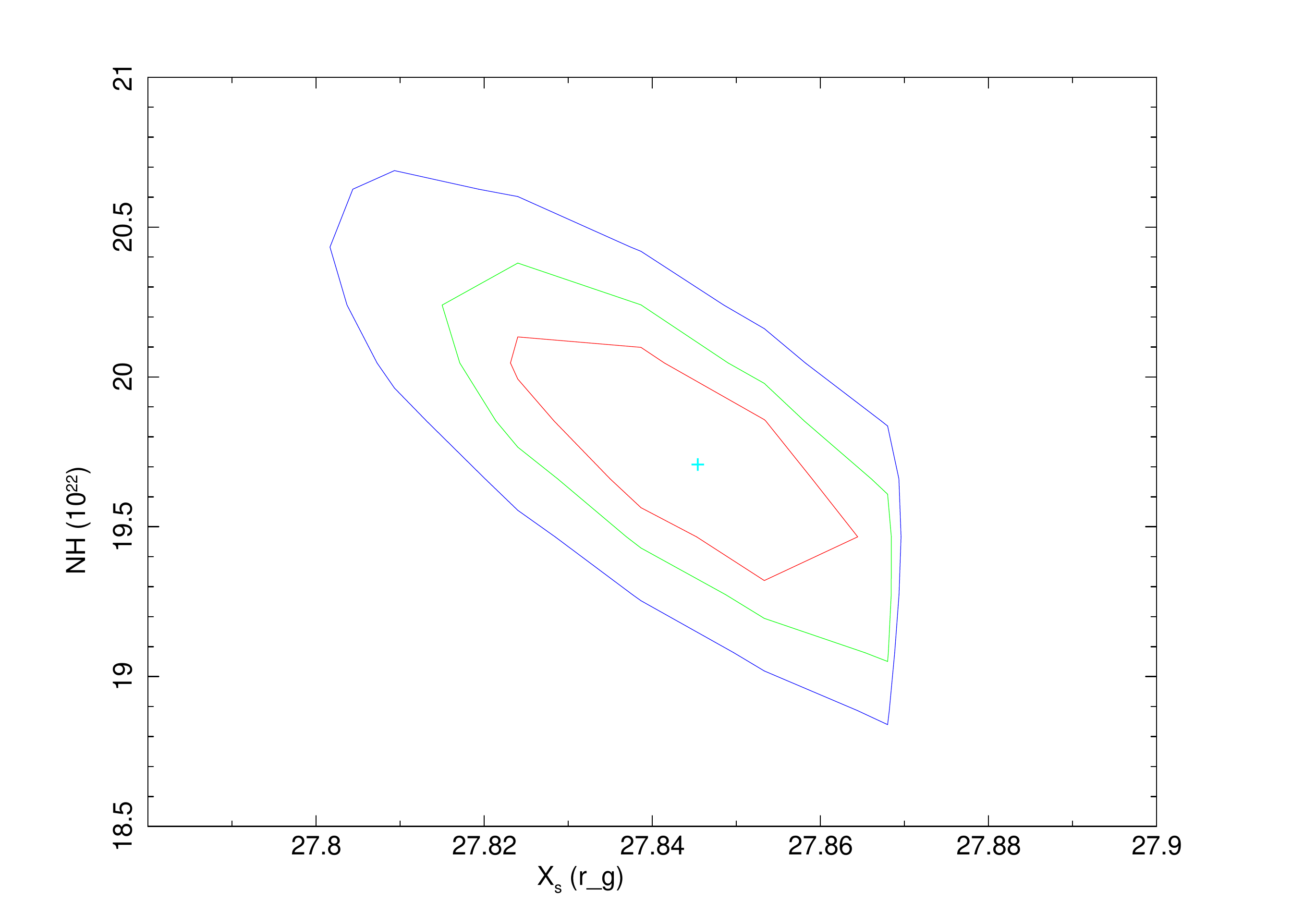}
\hspace{-1.0cm}
\includegraphics[height=6.5truecm,angle=270]{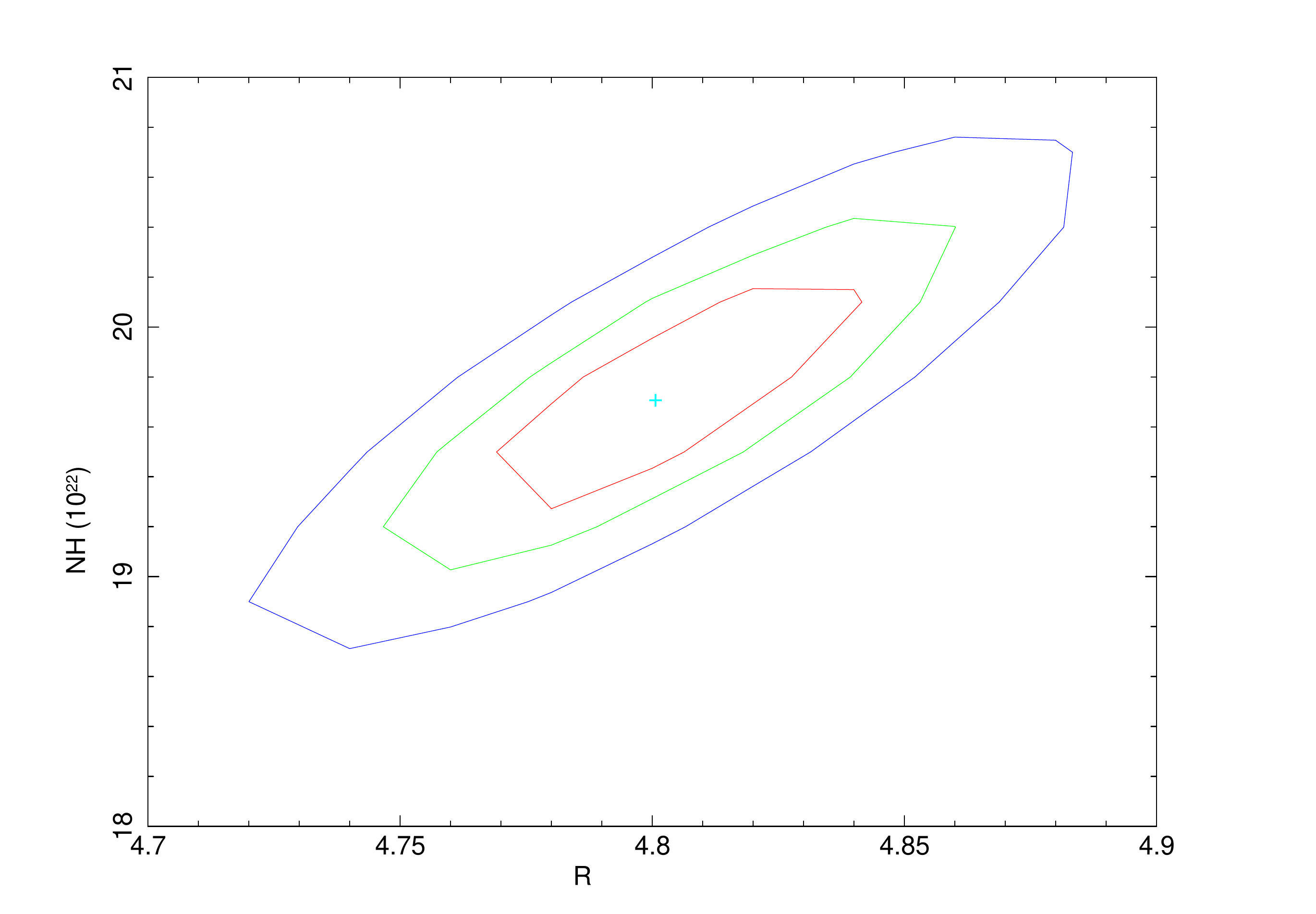}\hspace{-1.0cm}
\includegraphics[height=6.5truecm,angle=270]{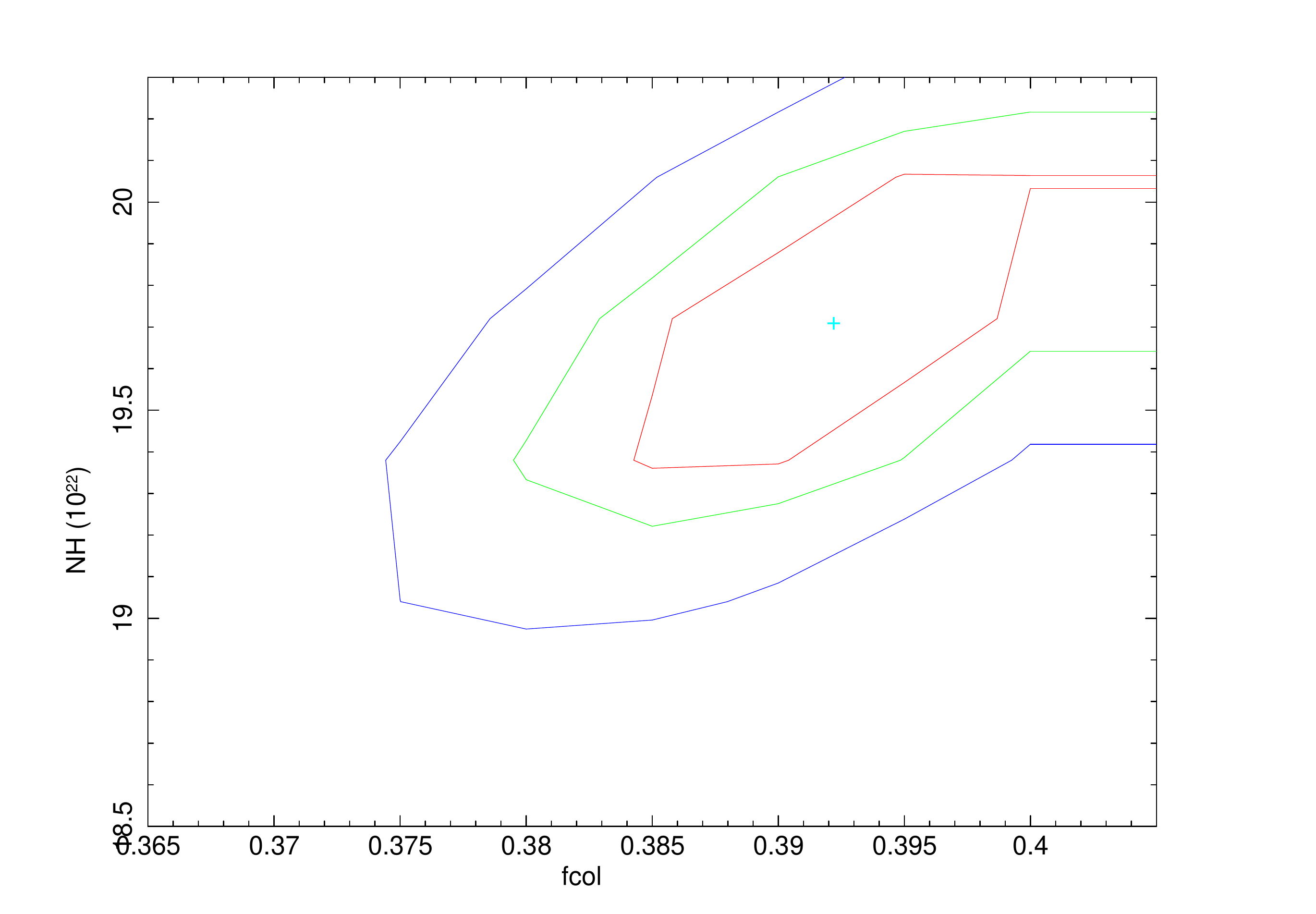}}
\caption{Confidence contour of $N_{H}$ with different continuum model parameters for the epoch A. Three different contour colors show the confidence level of one (red), two (green), and three sigma (blue). Here, x-labels follow the parameter notations used in \autoref{table:JetcafResults}, e.g. mdotd, mdoth, and fcol represents $\dot m_d$, $\dot m_h$, and $f_{col}$ respectively.}
\label{fig:ConfCont}
\end{figure*}

\autoref{fig:ConfCont} shows the confidence contours of $N_{H}$ with other \jetcaf\, model parameters. We have used {\it steppar} command in {\it XSPEC} to generate these plots. Different model parameters are labelled in x-axis. Three different contour colors (red, green, and blue) correspond to one, two, and three sigma confidence level. It is also the same as $\Delta \chi^2$ fit statistics of 2.3, 4.61, and 9.21. The correlation of $N_H$ with model parameters are clearly visible from the contour plots.  
The uncertainty of $N_H$ fits within 0.6 dex for 2 sigma confidence level. The other parameters assess better precision. The correlations are clear, which show that there is no obvious degeneracy between parameters and the solution is robust.

The top panel shows direct correlation of $N_H$, with M$_{BH}$, $\dot m_d$, and $\dot m_h$, implies that if these parameters value increase, continuum luminosity will also increase, therefore continuum driven wind will also contain more material along the LOS. However, in the case of $X_s$ is opposite, as it anti-correlates with the disc accretion rate. In the case of R \citep[up to the strong shock condition value, which is 4; see][]{Chakrabarti1999} and $f_{col}$, increase in both the parameters, increase the luminosity contribution from the jet, therefore the wind material launching along the LOS will increase. This gives a positive correlation. We note that in the case of, $f_{col}$ the contour is not closed at its higher value side (the upper limit of the grid value for this parameter is 0.5), this can be an artefact. Our continuum driven wind launching correlation study agrees with the work by \citet[][and references therein]{GoffordEtal2015}, where authors studied (anti)correlation of continuum luminosity with different physical quantities of the wind applicable for a large sample.

\begin{figure}
\includegraphics[height=6truecm]{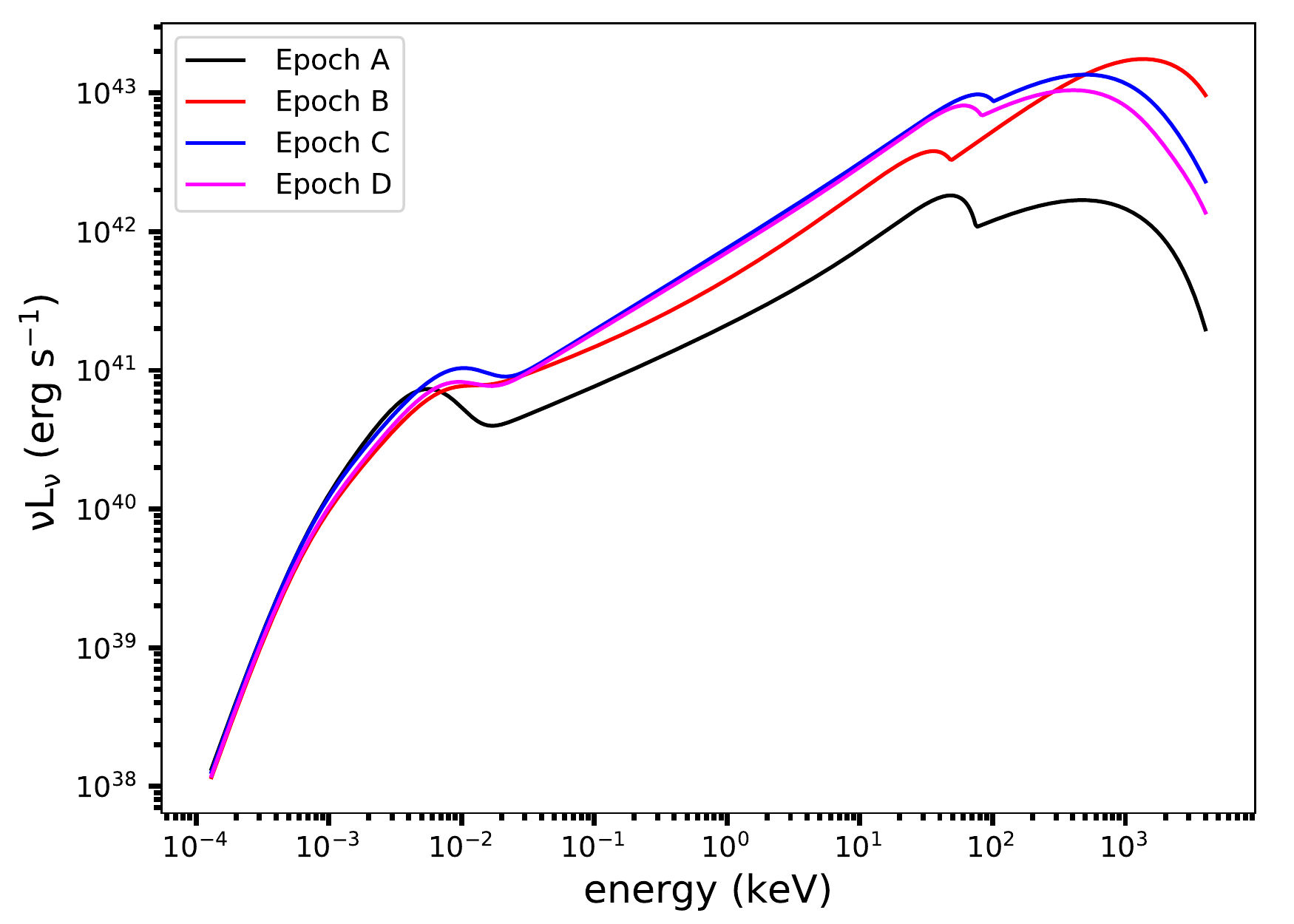}
\caption {The incident SEDs used in {\sc cloudy} models for the four epochs considered here. The SEDs are computed using the \jetcaf\,model which includes relevant physical processes in the accretion disc. The EUV flux is computed assuming the disc is emitting modified multi-color blackbody photons from each radius of the disc.}
\label{fig:Incid}
\end{figure}

As an additional test, we used photoionization simulation to predict absorption and emission line spectra for the modeled intrinsic spectral energy distributions (SEDs) shown in \autoref{fig:Incid}. The SED includes both disc (UV/optical photons), corona, and the jet component (hard X-ray) as discussed in \autoref{sec:Modelling}. In JeTCAF model, we compute the EUV photons considering the disc to be multi-temperature at each radius. At the zeroth iteration, the flux emitted from the disc is purely Keplerian \citep{Shakura1973}. These seed photons are getting upscattered by the hot corona. As the iterations go on, the interception of corona photons by the disc is taken into consideration, which modifies the original disc temperature.
In order to estimate the transmitted spectrum that an observer can measure along the LOS, we performed the numerical simulation of radiation transfer using the photoionization code {\sc cloudy}17.02 \citep{Ferland2017}. The essential parameters: the spectral energy distribution (SED), gas density $n_{\rm H}$, column density $N_{\rm H}$ are adopted in such a way that satisfies the spectral fitting from Table~\ref{table:JetcafResults}. The illuminating SEDs used in {\sc cloudy} are the \jetcaf ~model component from Fig.~\ref{fig:SpecFits} fitted for each epoch (excluding the other components). The gas density values $n_{\rm H}=2.65 \times 10^{10}$, $1.73 \times 10^{11}$,$1.47 \times 10^{11}$ and $1.01 \times 10^{11}$ cm$^{-3}$ are taken for the epochs A to D respectively.

Likewise, the column density values, $N_{\rm H}$= $1.97\times10^{23}$, $2.38\times10^{22}$, $0.8\times10^{22}$ and $4.13 \times 10^{22}$ cm$^{-2}$ are adopted from Column 8 of the Table~\ref{table:JetcafResults} for the epochs A to D respectively. The models are computed for the two representative cases of ionization parameter $\xi$=$10^{2}$ and $10^{3}$ erg cm s$^{-1}$ respectively. The $N_H$ and $\xi$ values fall within the range estimated in the literature \citep[][and references therein]{GoffordEtal2013,Waltonetal2014,BraitoEtal2014}. The $N_H$ values presently obtained are approximated values, as the fit does not consider ionization effects to the continuum. In all of these simulations, we considered the {\it Solar} values of the chemical composition, default in {\sc cloudy} adopted from \citet{Grevesse1998}. In order to account for the effects of micro-turbulence in absorption-emission lines, typical turbulent velocity $v_{\rm turb}=1000$ kms$^{-1}$ is used. Here we are not estimating or quantifying the turbulent velocity rather considering a typical value as a test case to generate the observed lines, however, for a detailed study for this effect we refer the work by \citet{GoffordEtal2013}. Note that, only the classical approach of one dimensional radiation transfer with plane-parallel open geometry without the inclusion of relativistic effects are considered. For the details of {\sc cloudy} applications in various AGN environments and generic definitions of parameters used, we refer the reader to the relevant literature \citep [for e.g.][]{Adhikari2015,Adhikari2019,Adhikari2019b} and cloudy documentation {\it Hazy} files\footnote{https://gitlab.nublado.org/cloudy/cloudy/-/wikis/home}.

\begin{figure}
\includegraphics[scale=0.6]{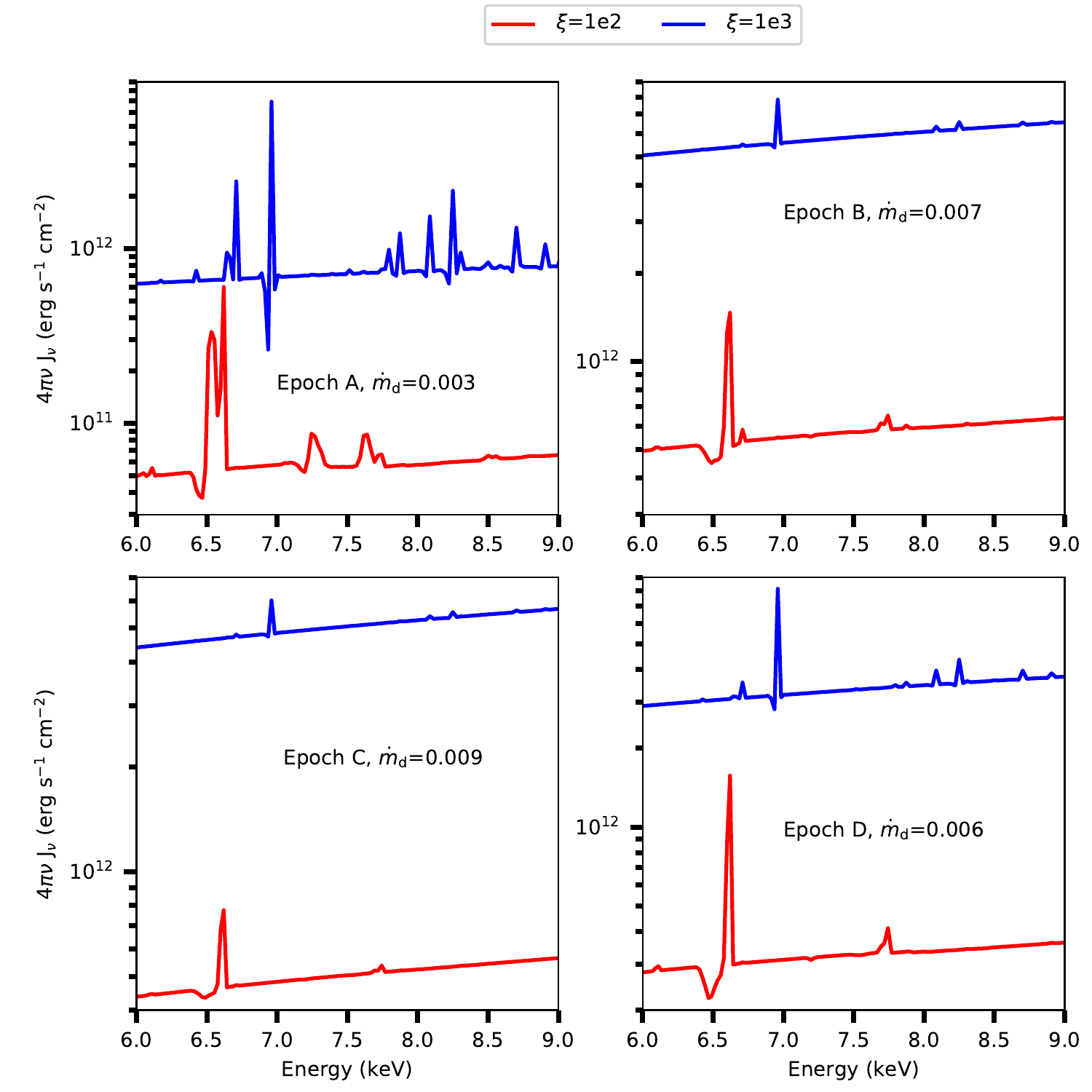}
\caption {The simulated spectra zoomed between $6$ to $9$ keV for four different epochs using {\sc cloudy} photoionization code. The evolution of the absorption properties with the change in the disc mass accretion rates (as seen from the fitted data) are plotted in each panel. The red and blue lines correspond to $\xi=$ 100 and 1000 respectively. The micro-turbulence of $v_{\rm turb}=1000$ km s$^{-1}$ is used in the simulations for all epochs.}
\label{fig:cloudy_spec}
\end{figure}

 The transmitted spectra zoomed in the band $6-9$ keV 
 are presented in the Fig.~\ref{fig:cloudy_spec}. The results from {\sc cloudy} simulations show absorption in the region around 6.4\,keV, which is similar to that seen in the observed spectra  
 shown in Fig.~\ref{fig:SpecFits}. The change in the spectral shapes, reflected in the illuminating SEDs in {\sc cloudy}, during the transition from epoch A to epoch D also changes the extent of Fe absorption. For the lowest value of mass accretion rate (0.003 $\dot M_{\rm Edd}$) in epoch A, there is a considerable amount of Fe absorption, and it becomes weak as the accretion rate becomes maximum (0.009 $\dot M_{\rm Edd}$) in epoch C. This result from basic simulations in {\sc cloudy} nicely corroborates with the observed features of absorption presented in Fig.~\ref{fig:SpecFits}. However, the emission lines observed are somehow skewed or broadened by the general relativistic (GR) effects of the highly spinning source. These effects are not included in {\sc cloudy} spectra and thus several components of narrow emission lines are present in Fig.~\ref{fig:cloudy_spec}.
 We noticed that the energy at which the absorption is significant depends on the adopted value of the $\xi$ in the photoionization models. The increase in $\xi$ allows for the ionization of the Fe to higher levels and as a result significant absorption towards the higher energy bands are seen. We note that the current CLOUDY modelling is to check whether the model SEDs for a given ionization parameter and turbulent velocity can generate emission and absorption lines similar to what is observed for NGC\, 1365 or not. However, the current approach does not measure the $\xi$ parameter robustly, it requires a detailed study. We plan to address this issue with grid of photoionization simulations with variations in ionization parameter in the future work.

\subsection{Failed wind scenario}
We consider here the case of the absorber is being associated with the outflowing matter. One of the most prominent outflowing region within AGNs is the broad line region (BLR), it is also one of the candidate to host intervening matter. Studies of warm absorbers (WA) observed at X-rays consider disc outflow or BLR as a source of absorption \citep[][]{Maiolino2010,Risaliti2011} or at least one of its component \citep[like NGC 5548 case described by][]{Kaastraetal2014}. Also, torus can be responsible for part of absorption \citep[Mrk 509 study by][]{Kaastraetal2012}. The expected ionization parameter value suggested here, ${\rm log}\ \xi \sim 2-3$, agrees with the values expected at the BLR position. Thus, we give here theoretical consideration of the BLR-like outflow. One of the possible outflow scenario is the failed radiatively driven outflow (FRADO) model, which clearly connects the BLR with the underlying accretion disc \citep{CzernyHryniewicz2011}.

We assume that the wind moves away from the disc with constant speed $v_{\rm w}$, 
in the form of a partial thin spherical shell at a distance $R_{\rm BLR}$ from the central source, the amount of mass per unit time carried out by the wind is:
\begin{equation}
    \dot M_{\rm w}=\mu m_p N_{\rm H} v_{\rm w} R_{\rm BLR} \Omega, 
\end{equation}
where, $\mu=1.4$, $m_p=1.67\times 10^{-24}$~gm, $\Omega=\pi/2$ \citep{Maiolino1995}, and 
$R_{\rm BLR}$ are the mean atomic mass per proton, proton mass, typical value of the solid angle subtended by the outflow, and the radius of the BLR respectively. From our spectral fitting, we found that
$N_{\rm H}$ decreases with increasing $\dot m_{\rm d}$, follows the relation $\mathcal{N} \dot m_{\rm d}^{-\beta}$. For NGC\,1365, $\mathcal{N} \sim 1.5 \times 10^{17}$ cm$^{-2}$ and $\beta$, $2.43\pm0.15$. 

To estimate the radial position of the outflowing matter we used an empirical relation based on reverberation mapping results for nearby AGNs \citep[ie.][]{Bentzetal2009} which is supported and explained by the FRADO model.
The BLR radius can be estimated from \citet{CzernyHryniewicz2011},

\begin{equation} \label{eq:rblrVsL}
    \log R_{\rm BLR}=1.538\pm0.027 + 0.5\log L_{44, 5100}
\end{equation}

where $R_{\rm BLR}$ is in light days and $L_{44,5100}$ is the monochromatic luminosity at 5100 \AA ~($\lambda L_{\lambda}$) measured in units of $10^{44}$ erg s$^{-1}$.
We can translate $L_{44,5100}$ to the bolometric luminosity using the relation
 $L_{\rm bol}=10.33 L_{\rm 5100}$ \citep{Richardsetal2006}. The $L_{\rm bol}$ is calculated using the relation $\eta \dot m_d c^2$, where, $\eta=0.1$ is considered. Our estimated values for $R_{\rm BLR}$ (in cm) using \autoref{eq:rblrVsL} are, \{$3.61\times10^{15}$, $5.47\times10^{15}$, $6.27\times10^{15}$, and $5.14\times10^{15}$\} for the epochs A to D are consistent with the literature. 

As the radiation from the central region falls onto the wind, 
and considering that on average each emitted photon scatters about 
once before escaping to infinity, this suggests that the total 
wind momentum must be of the order of the photon momentum
\citep{King2010} 
can be written as
\begin{equation}
    \dot M_{\rm w} v_{\rm w} \simeq \frac{L_{\rm bol}}{c}, 
\end{equation}
here $c$ is the speed of light.

After solving few steps, we find the equation for the wind velocity as,

\begin{equation}
    v_{\rm w}=\frac{L_{\rm bol}^{0.5}\dot m_{\rm d}^{\beta/2}}{(\mathcal{N} c m_p R_{\rm BLR} \mu \Omega)^{1/2}}=30 \left(\frac{L_{\rm bol}\dot m_{\rm d}^\beta}{\mathcal{N} R_{\rm BLR}}\right)^{1/2} ~~\text{km}~~\text{s}^{-1} .
\end{equation}

Our estimated wind velocities for all four epochs from A to D are: 1445, 4978, 7235, and 4001 km s$^{-1}$ respectively. The estimated wind velocities fall in the range reported in earlier studies on continuum driven wind launching \citep[see][and references therein]{GoffordEtal2013,BraitoEtal2014,GoffordEtal2015}.
In epochs A and D, the velocity values are less than the wind terminal velocity attained maximum (5748 and 4839 km~~s$^{-1}$ estimated using $\sqrt{2 G M_{\rm BH}/R_{\rm BLR}}$, here, G = Gravitational constant). For the epoch B, wind velocity is comparable with its terminal velocity (4623 km~~s$^{-1}$) of the wind. This failed wind behaves as a cloud shielding the central emission. However, in epoch C wind velocity higher than the terminal velocity (4365 km~~s$^{-1}$) can help to drive the material away from the disc, therefore, much less column density was observed. A similar scenario was studied earlier for a low mass black hole binaries \citep{Milleretal2020}.

From the model fits, the high absorption; variable column density along the LOS, which we argue could be due to the 
failed wind above the disc surface. The disc accretion rate was low, implies the radiative pressure was also low, therefore the
wind launched from the disc could stay in place and the matter was not pushed by the radiation to outer radii. In addition, strong ionizing radiation or outflows (which often is observed in the hard or intermediate spectral states, when accretion rate is low) from the central region of the disc itself can over ionize the wind from the disc, that can prevent the wind being driven away from the disc, therefore its velocity decreases \citep[for an extensive study in][]{Progaetal2004}. On the other hand, the failed wind BLR model suggests that the outflow velocity scales with the accretion rate. This happens because of its strong connection with the underlying accretion disc. As the dusty matter is pushed away from the disc atmosphere by the radiation from the underlying disc and later is exposed to the central radiation, which breaks its driving force or  pushes it away stronger to the outer radii \citep[][]{Naddaf-Bozena2021}. There are other physical mechanisms that can produce different emission/absorption lines in AGN environment, recently discussed in \citet[][for a review]{Lahaetal2021}.

\section{Conclusions} \label{sec:Conclusion}

In this paper, we study the changing look active galactic nucleus NGC\,1365 using accretion-ejection model including absorption and line emission in order to understand the maximum possible underlying physical processes behind this changing look
phenomenon. From our analysis, we conclude that:
\begin{itemize}
\item The accretion rates and their ratio, and the size of the dynamic corona varied significantly during the observation period.
\item The compression ratio of the shock at the boundary layer of the dynamic corona lies between 3.9 to 6.2, which is used to estimate the mass outflow rate from the dynamic corona. The maximum outflow rate estimated is $\sim 12\%$ of the inflow rate. This confirms the previous findings that the outflows are present in the system.
\item The geometry/size of the dynamic corona is also responsible for variable luminosity, that correlates with the hydrogen column density along the line of sight.
\item The accretion-ejection model fitted mass is placed within a range between $4.38\pm0.34 - 4.51\pm 0.29 \times 10^{6} M_\odot$, remarkably matches with the estimation by \citet{Onorietal2017}.
\item The presence of prominent absorption line and high hydrogen column density is due to low mass accretion rate. The low accretion rate is not able to generate enough pressure to push the matter farther out from the disc \citep[consistent with recent simulation by][]{Naddaf-Bozena2021}, therefore, it blocks the radiation coming from the inner region. This is also evidenced from the {\sc cloudy} simulations, where absorption features are stronger (at the highest $N_{\rm H}$) for the low mass accretion state. This explains the ``failed wind" scenario due to low wind velocity. 
\item We studied the correlation between $N_H$ with other model fitted continuum parameters (see \autoref{sec:Contour}) and found that they are consistent with the physical picture of the model and agree with the earlier works on continuum driven wind launching \citep[][and references therein]{GoffordEtal2015}.
\item Therefore, CLAGNs can show significant variability if the accretion rate and the corona geometry change substantially.

\end{itemize}    

\begin{acknowledgements}
We thank the referee for helpful comments and suggestions that improved the quality of the manuscript. SM acknowledges funding from Ramanujan Fellowship grant (\# RJF/2020/000113) by SERB-DST, Govt. of India. T. P. A gratefully  acknowledges the Inter-University Center for Astronomy and Astrophysics (IUCAA), Pune, India for providing the access to the Computational Cluster, where the numerical simulations used in this paper are performed.
This research has made use of the {\it NuSTAR} Data Analysis Software ({\sc nustardas}) jointly developed by the ASI Science Data Center (ASDC), Italy and the California Institute of Technology (Caltech), USA. This research has also made use of data obtained through the High Energy Astrophysics Science Archive Research Center Online Service, provided by NASA/Goddard Space Flight Center.

\end{acknowledgements}

\bibliography{clagns}{}

\begin{thebibliography}{76}
\expandafter\ifx\csname natexlab\endcsname\relax\def\natexlab#1{#1}\fi

\bibitem[{{Adhikari}(2019)}]{Adhikari2019b}
{Adhikari}, T.~P. 2019, {Photoionization Modelling as a Density Diagnostic of
  Line Emitting/Absorbing Regions in Active Galactic Nuclei}

\bibitem[{{Adhikari} {et~al.}(2019){Adhikari}, {R{\'o}{\.z}a{\'n}ska},
  {Hryniewicz}, {Czerny}, \& {Behar}}]{Adhikari2019}
{Adhikari}, T.~P., {R{\'o}{\.z}a{\'n}ska}, A., {Hryniewicz}, K., {Czerny}, B.,
  \& {Behar}, E. 2019, \apj, 881, 78

\bibitem[{{Adhikari} {et~al.}(2015){Adhikari}, {R{\'o}{\.z}a{\'n}ska},
  {Sobolewska}, \& {Czerny}}]{Adhikari2015}
{Adhikari}, T.~P., {R{\'o}{\.z}a{\'n}ska}, A., {Sobolewska}, M., \& {Czerny},
  B. 2015, \apj, 815, 83

\bibitem[{{Arnaud}(1996)}]{Arnaud1996}
{Arnaud}, K.~A. 1996, Astronomical Society of the Pacific Conference Series,
  Vol. 101, {XSPEC: The First Ten Years}, ed. G.~H. {Jacoby} \& J.~{Barnes}, 17

\bibitem[{{Balokovi{\'c}} {et~al.}(2018){Balokovi{\'c}}, {Brightman},
  {Harrison}, {Comastri}, {Ricci}, {Buchner}, {Gandhi}, {Farrah}, \&
  {Stern}}]{MislavEtal2018}
{Balokovi{\'c}}, M., {Brightman}, M., {Harrison}, F.~A., {et~al.} 2018, \apj,
  854, 42

\bibitem[{{Bentz} {et~al.}(2009){Bentz}, {Peterson}, {Netzer}, {Pogge}, \&
  {Vestergaard}}]{Bentzetal2009}
{Bentz}, M.~C., {Peterson}, B.~M., {Netzer}, H., {Pogge}, R.~W., \&
  {Vestergaard}, M. 2009, \apj, 697, 160

\bibitem[{{Braito} {et~al.}(2014){Braito}, {Reeves}, {Gofford}, {Nardini},
  {Porquet}, \& {Risaliti}}]{BraitoEtal2014}
{Braito}, V., {Reeves}, J.~N., {Gofford}, J., {et~al.} 2014, \apj, 795, 87

\bibitem[{{Brenneman} {et~al.}(2013){Brenneman}, {Risaliti}, {Elvis}, \&
  {Nardini}}]{BrennemanEtal2013}
{Brenneman}, L.~W., {Risaliti}, G., {Elvis}, M., \& {Nardini}, E. 2013, \mnras,
  429, 2662

\bibitem[{{Chakrabarti} \& {Titarchuk}(1995)}]{ChakrabartiTitarchuk1995}
{Chakrabarti}, S. \& {Titarchuk}, L.~G. 1995, \apj, 455, 623

\bibitem[{{Chakrabarti}(1989)}]{Chakrabarti1989}
{Chakrabarti}, S.~K. 1989, \apj, 347, 365

\bibitem[{{Chakrabarti}(1999)}]{Chakrabarti1999}
{Chakrabarti}, S.~K. 1999, \aap, 351, 185

\bibitem[{{Combes} {et~al.}(2019){Combes}, {Garc{\'\i}a-Burillo}, {Audibert},
  {Hunt}, {Eckart}, {Aalto}, {Casasola}, {Boone}, {Krips}, {Viti}, {Sakamoto},
  {Muller}, {Dasyra}, {van der Werf}, \& {Martin}}]{CombesEtal2019}
{Combes}, F., {Garc{\'\i}a-Burillo}, S., {Audibert}, A., {et~al.} 2019, \aap,
  623, A79

\bibitem[{{Connolly} {et~al.}(2014){Connolly}, {McHardy}, \&
  {Dwelly}}]{ConnolyEtal2014}
{Connolly}, S.~D., {McHardy}, I.~M., \& {Dwelly}, T. 2014, \mnras, 440, 3503

\bibitem[{{Czerny} \& {Hryniewicz}(2011)}]{CzernyHryniewicz2011}
{Czerny}, B. \& {Hryniewicz}, K. 2011, \aap, 525, L8

\bibitem[{{Debnath} {et~al.}(2014){Debnath}, {Chakrabarti}, \&
  {Mondal}}]{Debnathetal2014}
{Debnath}, D., {Chakrabarti}, S.~K., \& {Mondal}, S. 2014, \mnras, 440, L121

\bibitem[{{Denney} {et~al.}(2014){Denney}, {De Rosa}, {Croxall}, {Gupta},
  {Bentz}, {Fausnaugh}, {Grier}, {Martini}, {Mathur}, {Peterson}, {Pogge}, \&
  {Shappee}}]{Denneyetal2014}
{Denney}, K.~D., {De Rosa}, G., {Croxall}, K., {et~al.} 2014, \apj, 796, 134

\bibitem[{{Edmunds} {et~al.}(1988){Edmunds}, {Taylor}, \&
  {Turtle}}]{Edmundsetal1988}
{Edmunds}, M.~G., {Taylor}, K., \& {Turtle}, A.~J. 1988, \mnras, 234, 155

\bibitem[{{Fazeli} {et~al.}(2019){Fazeli}, {Busch}, {Valencia-S.}, {Eckart},
  {Zaja{\v{c}}ek}, {Combes}, \& {Garc{\'\i}a-Burillo}}]{Fazelietal2019}
{Fazeli}, N., {Busch}, G., {Valencia-S.}, M., {et~al.} 2019, \aap, 622, A128

\bibitem[{{Ferland} {et~al.}(2017){Ferland}, {Chatzikos}, {Guzm{\'a}n},
  {Lykins}, {van Hoof}, {Williams}, {Abel}, {Badnell}, {Keenan}, {Porter}, \&
  {Stancil}}]{Ferland2017}
{Ferland}, G.~J., {Chatzikos}, M., {Guzm{\'a}n}, F., {et~al.} 2017, \rmxaa, 53,
  385

\bibitem[{{Gao} {et~al.}(2021){Gao}, {Egusa}, {Liu}, {Kohno}, {Bao},
  {Morokuma-Matsui}, {Kong}, \& {Chen}}]{Gaoetal2021}
{Gao}, Y., {Egusa}, F., {Liu}, G., {et~al.} 2021, \apj, 913, 139

\bibitem[{{Gofford} {et~al.}(2015){Gofford}, {Reeves}, {McLaughlin}, {Braito},
  {Turner}, {Tombesi}, \& {Cappi}}]{GoffordEtal2015}
{Gofford}, J., {Reeves}, J.~N., {McLaughlin}, D.~E., {et~al.} 2015, \mnras,
  451, 4169

\bibitem[{{Gofford} {et~al.}(2013){Gofford}, {Reeves}, {Tombesi}, {Braito},
  {Turner}, {Miller}, \& {Cappi}}]{GoffordEtal2013}
{Gofford}, J., {Reeves}, J.~N., {Tombesi}, F., {et~al.} 2013, \mnras, 430, 60

\bibitem[{{Grevesse} \& {Sauval}(1998)}]{Grevesse1998}
{Grevesse}, N. \& {Sauval}, A.~J. 1998, \ssr, 85, 161

\bibitem[{{Guainazzi} {et~al.}(2009){Guainazzi}, {Risaliti}, {Nucita}, {Wang},
  {Bianchi}, {Soria}, \& {Zezas}}]{Guainazzietal2009}
{Guainazzi}, M., {Risaliti}, G., {Nucita}, A., {et~al.} 2009, \aap, 505, 589

\bibitem[{{Harrison} {et~al.}(2013){Harrison}, {Craig}, {Christensen},
  {Hailey}, {Zhang}, {Boggs}, {Stern}, {Cook}, {Forster}, {Giommi},
  {Grefenstette}, {Kim}, {Kitaguchi}, {Koglin}, {Madsen}, {Mao}, {Miyasaka},
  {Mori}, {Perri}, {Pivovaroff}, {Puccetti}, {Rana}, {Westergaard}, {Willis},
  {Zoglauer}, {An}, {Bachetti}, {Barri{\`e}re}, {Bellm}, {Bhalerao},
  {Brejnholt}, {Fuerst}, {Liebe}, {Markwardt}, {Nynka}, {Vogel}, {Walton},
  {Wik}, {Alexander}, {Cominsky}, {Hornschemeier}, {Hornstrup}, {Kaspi},
  {Madejski}, {Matt}, {Molendi}, {Smith}, {Tomsick}, {Ajello}, {Ballantyne},
  {Balokovi{\'c}}, {Barret}, {Bauer}, {Blandford}, {Brandt}, {Brenneman},
  {Chiang}, {Chakrabarty}, {Chenevez}, {Comastri}, {Dufour}, {Elvis}, {Fabian},
  {Farrah}, {Fryer}, {Gotthelf}, {Grindlay}, {Helfand}, {Krivonos}, {Meier},
  {Miller}, {Natalucci}, {Ogle}, {Ofek}, {Ptak}, {Reynolds}, {Rigby},
  {Tagliaferri}, {Thorsett}, {Treister}, \& {Urry}}]{Harrisonetal2013}
{Harrison}, F.~A., {Craig}, W.~W., {Christensen}, F.~E., {et~al.} 2013, \apj,
  770, 103

\bibitem[{{Hutsem{\'e}kers} {et~al.}(2019){Hutsem{\'e}kers}, {Ag{\'\i}s
  Gonz{\'a}lez}, {Marin}, {Sluse}, {Ramos Almeida}, \& {Acosta
  Pulido}}]{Hutsemekers2019}
{Hutsem{\'e}kers}, D., {Ag{\'\i}s Gonz{\'a}lez}, B., {Marin}, F., {et~al.}
  2019, \aap, 625, A54

\bibitem[{{Kaastra} {et~al.}(2012){Kaastra}, {Detmers}, {Mehdipour}, {Arav},
  {Behar}, {Bianchi}, {Branduardi-Raymont}, {Cappi}, {Costantini}, {Ebrero},
  {Kriss}, {Paltani}, {Petrucci}, {Pinto}, {Ponti}, {Steenbrugge}, \& {de
  Vries}}]{Kaastraetal2012}
{Kaastra}, J.~S., {Detmers}, R.~G., {Mehdipour}, M., {et~al.} 2012, \aap, 539,
  A117

\bibitem[{{Kaastra} {et~al.}(2014){Kaastra}, {Kriss}, {Cappi}, {Mehdipour},
  {Petrucci}, {Steenbrugge}, {Arav}, {Behar}, {Bianchi}, {Boissay},
  {Branduardi-Raymont}, {Chamberlain}, {Costantini}, {Ely}, {Ebrero}, {Di
  Gesu}, {Harrison}, {Kaspi}, {Malzac}, {De Marco}, {Matt}, {Nandra},
  {Paltani}, {Person}, {Peterson}, {Pinto}, {Ponti}, {Nu{\~n}ez}, {De Rosa},
  {Seta}, {Ursini}, {de Vries}, {Walton}, \& {Whewell}}]{Kaastraetal2014}
{Kaastra}, J.~S., {Kriss}, G.~A., {Cappi}, M., {et~al.} 2014, Science, 345, 64

\bibitem[{{Kalberla} {et~al.}(2005){Kalberla}, {Burton}, {Hartmann}, {Arnal},
  {Bajaja}, {Morras}, \& {P{\"o}ppel}}]{Kalberlaetal2005}
{Kalberla}, P.~M.~W., {Burton}, W.~B., {Hartmann}, D., {et~al.} 2005, \aap,
  440, 775

\bibitem[{{Kara} {et~al.}(2015){Kara}, {Zoghbi}, {Marinucci}, {Walton},
  {Fabian}, {Risaliti}, {Boggs}, {Christensen}, {Fuerst}, {Hailey}, {Harrison},
  {Matt}, {Parker}, {Reynolds}, {Stern}, \& {Zhang}}]{Karaetal2015}
{Kara}, E., {Zoghbi}, A., {Marinucci}, A., {et~al.} 2015, \mnras, 446, 737

\bibitem[{{Kim} {et~al.}(2018){Kim}, {Yoon}, \& {Evans}}]{Kimetal2018}
{Kim}, D.~C., {Yoon}, I., \& {Evans}, A.~S. 2018, \apj, 861, 51

\bibitem[{{King}(2010)}]{King2010}
{King}, A.~R. 2010, \mnras, 402, 1516

\bibitem[{{Laha} {et~al.}(2021){Laha}, {Reynolds}, {Reeves}, {Kriss},
  {Guainazzi}, {Smith}, {Veilleux}, \& {Proga}}]{Lahaetal2021}
{Laha}, S., {Reynolds}, C.~S., {Reeves}, J., {et~al.} 2021, Nature Astronomy,
  5, 13

\bibitem[{{LaMassa} {et~al.}(2015){LaMassa}, {Cales}, {Moran}, {Myers},
  {Richards}, {Eracleous}, {Heckman}, {Gallo}, \& {Urry}}]{LaMassaetal2015}
{LaMassa}, S.~M., {Cales}, S., {Moran}, E.~C., {et~al.} 2015, \apj, 800, 144

\bibitem[{{Lindblad}(1999)}]{Lindbladreview1999}
{Lindblad}, P.~O. 1999, \aapr, 9, 221

\bibitem[{{Maiolino} \& {Rieke}(1995{\natexlab{a}})}]{Maiolino1995}
{Maiolino}, R. \& {Rieke}, G.~H. 1995{\natexlab{a}}, \apj, 454, 95

\bibitem[{{Maiolino} \& {Rieke}(1995{\natexlab{b}})}]{MaiolinoReike1995}
{Maiolino}, R. \& {Rieke}, G.~H. 1995{\natexlab{b}}, \apj, 454, 95

\bibitem[{{Maiolino} {et~al.}(2010){Maiolino}, {Risaliti}, {Salvati},
  {Pietrini}, {Torricelli-Ciamponi}, {Elvis}, {Fabbiano}, {Braito}, \&
  {Reeves}}]{Maiolino2010}
{Maiolino}, R., {Risaliti}, G., {Salvati}, M., {et~al.} 2010, \aap, 517, A47

\bibitem[{{Matt} {et~al.}(2003){Matt}, {Guainazzi}, \&
  {Maiolino}}]{MattEtal2003}
{Matt}, G., {Guainazzi}, M., \& {Maiolino}, R. 2003, \mnras, 342, 422

\bibitem[{{McElroy} {et~al.}(2016){McElroy}, {Husemann}, {Croom}, {Davis},
  {Bennert}, {Busch}, {Combes}, {Eckart}, {Perez-Torres}, {Powell},
  {Scharw{\"a}chter}, {Tremblay}, \& {Urrutia}}]{McElroyetal2016}
{McElroy}, R.~E., {Husemann}, B., {Croom}, S.~M., {et~al.} 2016, \aap, 593, L8

\bibitem[{{Mehdipour} {et~al.}(2021){Mehdipour}, {Kriss}, {Brenneman},
  {Costantini}, {Kaastra}, {Branduardi-Raymont}, {Di Gesu}, {Ebrero}, \&
  {Mao}}]{MehdipourEtal2021}
{Mehdipour}, M., {Kriss}, G.~A., {Brenneman}, L.~W., {et~al.} 2021, arXiv
  e-prints, arXiv:2112.06297

\bibitem[{{Miller} {et~al.}(2020){Miller}, {Zoghbi}, {Raymond}, {Balakrishnan},
  {Brenneman}, {Cackett}, {Draghis}, {Fabian}, {Gallo}, {Kaastra}, {Kallman},
  {Kammoun}, {Motta}, {Proga}, {Reynolds}, \& {Trueba}}]{Milleretal2020}
{Miller}, J.~M., {Zoghbi}, A., {Raymond}, J., {et~al.} 2020, \apj, 904, 30

\bibitem[{{Mondal} \& {Chakrabarti}(2019)}]{MondalChakrabarti2019}
{Mondal}, S. \& {Chakrabarti}, S.~K. 2019, \mnras, 483, 1178

\bibitem[{Mondal \& Chakrabarti(2021)}]{MondalChakrabarti2021}
Mondal, S. \& Chakrabarti, S.~K. 2021, The Astrophysical Journal, 920, 41

\bibitem[{{Mondal} {et~al.}(2014{\natexlab{a}}){Mondal}, {Chakrabarti}, \&
  {Debnath}}]{MondalEtal2014ApSS}
{Mondal}, S., {Chakrabarti}, S.~K., \& {Debnath}, D. 2014{\natexlab{a}}, \apss,
  353, 223

\bibitem[{{Mondal} {et~al.}(2014{\natexlab{b}}){Mondal}, {Debnath}, \&
  {Chakrabarti}}]{Mondaletal2014}
{Mondal}, S., {Debnath}, D., \& {Chakrabarti}, S.~K. 2014{\natexlab{b}}, \apj,
  786, 4

\bibitem[{{Mondal} \& {Stalin}(2021)}]{MondalStalin2021}
{Mondal}, S. \& {Stalin}, C.~S. 2021, Galaxies, 9, 21

\bibitem[{{Naddaf} {et~al.}(2021){Naddaf}, {Czerny}, \&
  {Szczerba}}]{Naddaf-Bozena2021}
{Naddaf}, M.-H., {Czerny}, B., \& {Szczerba}, R. 2021, arXiv e-prints,
  arXiv:2102.00336

\bibitem[{{Nandi} {et~al.}(2019){Nandi}, {Chakrabarti}, \&
  {Mondal}}]{Nandietal2019}
{Nandi}, P., {Chakrabarti}, S.~K., \& {Mondal}, S. 2019, \apj, 877, 65

\bibitem[{{Nardini} {et~al.}(2015){Nardini}, {Gofford}, {Reeves}, {Braito},
  {Risaliti}, \& {Costa}}]{NardiniEtal2015}
{Nardini}, E., {Gofford}, J., {Reeves}, J.~N., {et~al.} 2015, \mnras, 453, 2558

\bibitem[{{Noda} \& {Done}(2018)}]{NodaDone2018}
{Noda}, H. \& {Done}, C. 2018, \mnras, 480, 3898

\bibitem[{{Onori} {et~al.}(2017){Onori}, {Ricci}, {La Franca}, {Bianchi},
  {Bongiorno}, {Brusa}, {Fiore}, {Maiolino}, {Marconi}, {Sani}, \&
  {Vignali}}]{Onorietal2017}
{Onori}, F., {Ricci}, F., {La Franca}, F., {et~al.} 2017, \mnras, 468, L97

\bibitem[{{Proga} \& {Kallman}(2004)}]{Progaetal2004}
{Proga}, D. \& {Kallman}, T.~R. 2004, \apj, 616, 688

\bibitem[{{Ricci} {et~al.}(2016){Ricci}, {Bauer}, {Arevalo}, {Boggs}, {Brandt},
  {Christensen}, {Craig}, {Gandhi}, {Hailey}, {Harrison}, {Koss}, {Markwardt},
  {Stern}, {Treister}, \& {Zhang}}]{Riccietal2016}
{Ricci}, C., {Bauer}, F.~E., {Arevalo}, P., {et~al.} 2016, \apj, 820, 5

\bibitem[{{Ricci} {et~al.}(2020){Ricci}, {Kara}, {Loewenstein}, {Trakhtenbrot},
  {Arcavi}, {Remillard}, {Fabian}, {Gendreau}, {Arzoumanian}, {Li}, {Ho},
  {MacLeod}, {Cackett}, {Altamirano}, {Gandhi}, {Kosec}, {Pasham}, {Steiner},
  \& {Chan}}]{Riccietal2020}
{Ricci}, C., {Kara}, E., {Loewenstein}, M., {et~al.} 2020, \apjl, 898, L1

\bibitem[{{Richards} {et~al.}(2006){Richards}, {Lacy}, {Storrie-Lombardi},
  {Hall}, {Gallagher}, {Hines}, {Fan}, {Papovich}, {Vanden Berk}, {Trammell},
  {Schneider}, {Vestergaard}, {York}, {Jester}, {Anderson}, {Budav{\'a}ri}, \&
  {Szalay}}]{Richardsetal2006}
{Richards}, G.~T., {Lacy}, M., {Storrie-Lombardi}, L.~J., {et~al.} 2006, \apjs,
  166, 470

\bibitem[{{Risaliti} {et~al.}(2005){Risaliti}, {Elvis}, {Fabbiano}, {Baldi}, \&
  {Zezas}}]{Risalitietal2005}
{Risaliti}, G., {Elvis}, M., {Fabbiano}, G., {Baldi}, A., \& {Zezas}, A. 2005,
  \apjl, 623, L93

\bibitem[{{Risaliti} {et~al.}(2002){Risaliti}, {Elvis}, \&
  {Nicastro}}]{Risalitietal2002}
{Risaliti}, G., {Elvis}, M., \& {Nicastro}, F. 2002, \apj, 571, 234

\bibitem[{{Risaliti} {et~al.}(2013){Risaliti}, {Harrison}, {Madsen}, {Walton},
  {Boggs}, {Christensen}, {Craig}, {Grefenstette}, {Hailey}, {Nardini},
  {Stern}, \& {Zhang}}]{Risalitietal2013}
{Risaliti}, G., {Harrison}, F.~A., {Madsen}, K.~K., {et~al.} 2013, \nat, 494,
  449

\bibitem[{{Risaliti} {et~al.}(2009){Risaliti}, {Miniutti}, {Elvis}, {Fabbiano},
  {Salvati}, {Baldi}, {Braito}, {Bianchi}, {Matt}, {Reeves}, {Soria}, \&
  {Zezas}}]{Risalitietal2009}
{Risaliti}, G., {Miniutti}, G., {Elvis}, M., {et~al.} 2009, \apj, 696, 160

\bibitem[{{Risaliti} {et~al.}(2011){Risaliti}, {Nardini}, {Salvati}, {Elvis},
  {Fabbiano}, {Maiolino}, {Pietrini}, \& {Torricelli-Ciamponi}}]{Risaliti2011}
{Risaliti}, G., {Nardini}, E., {Salvati}, M., {et~al.} 2011, \mnras, 410, 1027

\bibitem[{{Rivers} {et~al.}(2015){Rivers}, {Risaliti}, {Walton}, {Harrison},
  {Ar{\'e}valo}, {Baur}, {Boggs}, {Brenneman}, {Brightman}, {Christensen},
  {Craig}, {F{\"u}rst}, {Hailey}, {Hickox}, {Marinucci}, {Reeves}, {Stern}, \&
  {Zhang}}]{Riversetal2015}
{Rivers}, E., {Risaliti}, G., {Walton}, D.~J., {et~al.} 2015, \apj, 804, 107

\bibitem[{{Sandqvist} {et~al.}(1995){Sandqvist}, {Joersaeter}, \&
  {Lindblad}}]{Sandqvistetal1995}
{Sandqvist}, A., {Joersaeter}, S., \& {Lindblad}, P.~O. 1995, \aap, 295, 585

\bibitem[{{Shakura} \& {Sunyaev}(1973)}]{Shakura1973}
{Shakura}, N.~I. \& {Sunyaev}, R.~A. 1973, \aap, 500, 33

\bibitem[{{Sharp} \& {Bland-Hawthorn}(2010)}]{SharpBlandHawthorn2010}
{Sharp}, R.~G. \& {Bland-Hawthorn}, J. 2010, \apj, 711, 818

\bibitem[{{Sheng} {et~al.}(2017){Sheng}, {Wang}, {Jiang}, {Yang}, {Yan}, {Dou},
  \& {Peng}}]{Shengetal2017}
{Sheng}, Z., {Wang}, T., {Jiang}, N., {et~al.} 2017, \apjl, 846, L7

\bibitem[{{{\'S}niegowska} \& {Czerny}(2019)}]{Sniegowska2019}
{{\'S}niegowska}, M. \& {Czerny}, B. 2019, arXiv e-prints, arXiv:1904.06767

\bibitem[{{Storchi-Bergmann} \& {Bonatto}(1991)}]{Storchi-Bergmann1991}
{Storchi-Bergmann}, T. \& {Bonatto}, C.~J. 1991, \mnras, 250, 138

\bibitem[{{Titarchuk} \& {Shrader}(2005)}]{TitarchukShrader2005}
{Titarchuk}, L. \& {Shrader}, C. 2005, \apj, 623, 362

\bibitem[{{Veilleux} {et~al.}(2003){Veilleux}, {Shopbell}, {Rupke},
  {Bland-Hawthorn}, \& {Cecil}}]{Veilleuxetal2003}
{Veilleux}, S., {Shopbell}, P.~L., {Rupke}, D.~S., {Bland-Hawthorn}, J., \&
  {Cecil}, G. 2003, \aj, 126, 2185

\bibitem[{{Venturi} {et~al.}(2018){Venturi}, {Nardini}, {Marconi}, {Carniani},
  {Mingozzi}, {Cresci}, {Mannucci}, {Risaliti}, {Maiolino}, {Balmaverde},
  {Bongiorno}, {Brusa}, {Capetti}, {Cicone}, {Ciroi}, {Feruglio}, {Fiore},
  {Gallazzi}, {La Franca}, {Mainieri}, {Matsuoka}, {Nagao}, {Perna},
  {Piconcelli}, {Sani}, {Tozzi}, \& {Zibetti}}]{Venturietal2018}
{Venturi}, G., {Nardini}, E., {Marconi}, A., {et~al.} 2018, \aap, 619, A74

\bibitem[{{Veron} {et~al.}(1980){Veron}, {Lindblad}, {Zuiderwijk}, {Veron}, \&
  {Adam}}]{Veronetal1980}
{Veron}, P., {Lindblad}, P.~O., {Zuiderwijk}, E.~J., {Veron}, M.~P., \& {Adam},
  G. 1980, \aap, 87, 245

\bibitem[{{Walton} {et~al.}(2014){Walton}, {Risaliti}, {Harrison}, {Fabian},
  {Miller}, {Arevalo}, {Ballantyne}, {Boggs}, {Brenneman}, {Christensen},
  {Craig}, {Elvis}, {Fuerst}, {Gandhi}, {Grefenstette}, {Hailey}, {Kara},
  {Luo}, {Madsen}, {Marinucci}, {Matt}, {Parker}, {Reynolds}, {Rivers}, {Ross},
  {Stern}, \& {Zhang}}]{Waltonetal2014}
{Walton}, D.~J., {Risaliti}, G., {Harrison}, F.~A., {et~al.} 2014, \apj, 788,
  76

\bibitem[{{Whewell} {et~al.}(2016){Whewell}, {Branduardi-Raymont}, \&
  {Page}}]{Whewelletal2016}
{Whewell}, M., {Branduardi-Raymont}, G., \& {Page}, M.~J. 2016, \aap, 595, A85

\bibitem[{{Wilms} {et~al.}(2000){Wilms}, {Allen}, \& {McCray}}]{Wilmsetal2000}
{Wilms}, J., {Allen}, A., \& {McCray}, R. 2000, \apj, 542, 914

\bibitem[{{Yang} {et~al.}(2018){Yang}, {Wu}, {Fan}, {Jiang}, {McGreer},
  {Shangguan}, {Yao}, {Wang}, {Joshi}, {Green}, {Wang}, {Feng}, {Fu}, {Yang},
  \& {Liu}}]{Yangetal2018}
{Yang}, Q., {Wu}, X.-B., {Fan}, X., {et~al.} 2018, \apj, 862, 109

\end{thebibliography}
\bibliographystyle{aa}


\begin{appendix}


\end{appendix}

\end{document}